\documentclass[11pt]{article}
\usepackage{enumerate}
\usepackage{natbib}
\usepackage{amsmath,amsthm,amssymb}
\usepackage{authblk}
\usepackage{graphicx,psfrag,epsf}
\usepackage{epstopdf}
\usepackage{epsfig}
\usepackage[toc]{appendix}
\usepackage{bbm}
\usepackage{enumitem}
\usepackage{color}
\usepackage{url} 

\RequirePackage[hyperindex,breaklinks,colorlinks,citecolor=blue,urlcolor=black]{hyperref}

\theoremstyle{plain}

\newtheorem{prop}{Proposition}

\def\T{{ \top }}

\newcommand{\x}{\mathbf{x}}

\renewcommand{\v}{\mathbf{v}}

\newcommand{\y}{\mathbf{y}}
\newcommand{\z}{\mathbf{z}}

\newcommand{\bbe}{\boldsymbol{\beta}}
\newcommand{\bdel}{\boldsymbol{\delta}}
\newcommand{\bgam}{\boldsymbol{\gamma}}
\newcommand{\beps}{\boldsymbol{\epsilon}}

\newcommand{\bphi}{\boldsymbol{\phi}}
\newcommand{\bthe}{\boldsymbol{\theta}}

\newcommand{\bet}{\boldsymbol{\eta}}

\newcommand{\B}{\mathbf{B}}
\newcommand{\bb}{\mathbf{b}}

\newcommand{\W}{\mathbf{W}}
\newcommand{\X}{\mathbf{X}}

\newcommand{\Z}{\mathbf{Z}}

\newcommand\name{MSBS }
\newcommand\namenospace{MSBS}

\newcommand{\blind}{0}

\begin{document}

\def\spacingset#1{\renewcommand{\baselinestretch}%
{#1}\small\normalsize} \spacingset{1}


\if0\blind
{
  \title{\bf Bayesian model selection in additive partial linear models via locally adaptive splines}
  \author[1,2]{Seonghyun Jeong}
  \author[1,2]{Taeyoung Park\thanks{Corresponding author: tpark@yonsei.ac.kr}} 
  \author[3]{David A. van Dyk} 
  \affil[1]{Department of Statistics and Data Science, Yonsei University, Seoul, Korea}
  \affil[2]{Department of Applied Statistics, Yonsei University, Seoul, Korea}
  \affil[3]{Statistics Section, Department of Mathematics, Imperial College London, London, UK}
  \maketitle
} \fi

\if1\blind
{
  \bigskip
  \bigskip
  \bigskip
  \begin{center}
    {\LARGE\bf Title}
\end{center}
  \medskip
} \fi

\begin{abstract}
We provide a flexible framework for selecting among a class of additive partial linear models that allows both linear and nonlinear additive components.  In practice, it is challenging to determine which additive components should be excluded from the model while simultaneously determining whether nonzero additive components should be represented as linear or non-linear components in the final model. 
In this paper, we propose a Bayesian model selection method that is facilitated by a carefully specified class of models, including the choice of a prior distribution and the nonparametric model used for the nonlinear additive components. 
We employ a series of latent variables that determine the effect of each variable among the three possibilities (no effect, linear effect, and nonlinear effect) and that simultaneously determine the knots of each spline for a suitable penalization of smooth functions. The use of a pseudo-prior distribution along with a collapsing scheme enables us to deploy well-behaved Markov chain Monte Carlo samplers, both for model selection and for fitting the preferred model. Our method and algorithm are deployed on a suite of numerical studies and are applied to a nutritional epidemiology study. The numerical results show that the proposed methodology outperforms previously available methods in terms of effective sample sizes of the Markov chain samplers and the overall misclassification rates. 

\end{abstract}

\noindent%

\spacingset{1.1}

\section{Introduction}
\label{sec:intro}

An additive partial linear model (APLM) generalizes both a standard linear model and a nonparametric additive model by combining both linear and nonlinear additive components in the framework of an additive regression model. Let $Y\in\mathbb R$ be a response variable, let ${\cal X}=(X_1,\dots,X_p)^\T \in \mathbb{R}^p$ be a $p$-dimensional continuous predictor vector, and let ${\cal Z}=(Z_1,\dots,Z_q)^\T \in \mathbb{R}^q$ be a $q$-dimensional predictor vector that may be either continuous or discrete. An APLM is written as
\begin{align}
	Y ~=~ \alpha +\sum_{j=1}^p  f_j(X_j) + \sum_{k=1}^q \beta_k Z_k  + \epsilon,
	\label{eqn:aplm}
\end{align}
where $\alpha$ is an intercept term, $f_1,\dots,f_p$ are unknown functions corresponding to $X_1,\dots,X_p$, $\beta_1,\dots,\beta_q$ are coefficients corresponding to $Z_1,\dots,Z_q$, and $\epsilon$ is a Gaussian random error independent of the predictors $({\cal X}, {\cal Z})$ with mean zero and finite variance $\sigma^2$. The identifiability of $f_1,\dots,f_p$ in~(\ref{eqn:aplm}) is assured by the restriction ${\rm E}[f_j(X_j)]=0$ for $j=1,\dots,p$ \citep{xue:09,huan:etal:10}. The APLM in~(\ref{eqn:aplm}) is a useful compromise between the simplicity of a linear model and the flexibility of a nonparametric additive model, especially when the model involves categorical predictors. (A fully nonparametric additive structure for dummy variables associated with the categorical predictors reduces to a linear model.) Suppose that some additive components, $f_j$, are zero, linear, or nonlinear, while some coefficients, $\beta_k$, are zero or not. In this paper, we consider model selection for the APLM in~(\ref{eqn:aplm}), where nonzero additive components are distinguished from zero components  while it is simultaneously determined whether each $f_j$ can be further simplified as a strictly linear function via locally adaptive estimation.

Whereas there is a rich literature on both frequentist and Bayesian model selection methods that identify zero and nonlinear effects for the additive components in nonparametric additive models \citep{lin:zhan:06,ravi:etal:09,xue:09,huan:etal:10,rask:etal:12,curt:etal:14,bane:ghos:14},
the more general idea, which simultaneously identifies zero, linear, and nonlinear effects, has been sparsely considered. Frequentist approaches to this general problem typically involve many tuning parameters or multi-stage tuning procedures, which is challenging in practice \citep{lian:etal:15,wu:stef:15}. On the other hand, Bayesian approaches provide a more natural solution that involves the comparison of posterior probabilities between competing models, but the problem of controlling the smoothness of unknown functions in the nonlinear additive components remains. Taking a Bayesian perspective, \citet{ross:rubi:19} addressed the above-mentioned general model selection problem for survival models by introducing ternary latent variables. Although they provide a solid theoretical understanding of their method, smooth function estimation via an appropriate penalization is not considered in their work. \citet{sche:etal:11} and \citet{bove:etal:15} dealt with the model selection problem together with smooth function estimation, but their approaches to the model selection are less intuitive, and their methods for function estimation do not catch the locally varying smoothness (see the next paragraph for more details). This study aims to address these issues.

It still remains to discuss how the nonparametric additive components can be estimated simultaneously with model selection.
Several Bayesian strategies for estimating smooth functions have been developed, including Gaussian process regression, kernel regression, and spline regression.
Here we discuss two popular spline-based methods.
First, Bayesian penalized splines can be used to estimate unknown functions with fixed knot locations and a prior distribution on the smoothing parameters \citep{rupp:etal:03,lang:brez:04}. 
For general model selection in the APLM that we consider, 
\citet{sche:etal:11} and \citet{bove:etal:15} adopted the Bayesian penalized splines to estimate unknown functions in nonlinear additive components. However, this approach cannot capture the locally varying curvature of smooth functions without significant modifications \citep{rupp:etal:03}. Although several extensions to local adaptation have been proposed \citep{bala:etal:2005,jull:lamb:07,sche:knei:09}, they are not readily available for model selection in an APLM because they require additional tuning parameters that complicate the calculations needed for model comparison. Second, Bayesian regression splines can be used to estimate unknown functions with basis (or knot) selection \citep{smit:kohn:96,deni:etal:98,kohn:etal:01,dima:etal:01}. By considering candidates of possible basis functions (depending on knot locations) and selecting optimal ones based on data, unknown functions can be estimated with locally adaptive smoothing and without overfitting, thereby overcoming the primary challenge of the Bayesian penalized splines. 

An even more important advantage of the basis-selection approach is that model selection is accomplished as an automatic byproduct of the estimation of the nonlinear additive components in the APLM.
Specifically, if no basis functions are selected in the regression spline, the corresponding nonlinear additive component is zero, whereas if only a linear basis function is selected, then it is linear. Despite this apparent advantage, the basis-selection approach has not been widely used for identifying function classes and selecting the set of nonlinear additive components to include in an APLM in this way. This is because doing so would require simultaneously selecting basis functions for each regression spline of a possibly large number of nonlinear additive components, leading to time-consuming computational challenges \citep{sche:etal:11}.  The primary contribution of this paper is to show how a careful choice of prior distributions combined with an efficient Markov chain Monte Carlo (MCMC) posterior sampling technique allows us to deploy the basis-selection approach for model selection.

We refer to our proposed method as Model Selection via Basis Selection (\namenospace) in APLMs.
\name allows the user to specify which variables can be considered only as linear predictors and which can be considered as either linear or nonlinear.
It then automatically identifies predictors with zero or linear effects among the former and automatically identifies zero, linear, and nonlinear effects among the latter group. This, for example, allows dummy variables to be introduced without being confused as possible nonlinear predictors. We adopt a data-driven strategy to control both the number of basis functions and to select which basis functions to include from a large set. This allows us to flexibly and adaptively estimate the smooth predictor functions that additively enter the response function of the APLM.

Similar to \citet{ross:rubi:19}, we introduce a set of latent indicator variables that identify which predictor variables have
no effect, which have linear effects, and which have nonlinear effects.
Another set of latent variables allows for locally adaptive smoothness of the predictor functions via their basis selection. The proposed model employs semi-conjugate prior distributions such as Zellner's $g$-prior distribution \citep{zell:86}, which allows us to analytically compute the marginal posterior distributions of certain subsets of the parameters. By iteratively sampling from the conditional distributions of either the joint posterior distribution or its marginal posterior distribution, we construct an efficient partially-collapsed Gibbs sampler \citep{vand:park:08}. Convergence of the sampler is further improved via the ``pseudo-prior'' method developed by \citet{carl:chib:95}.  Finally, we introduce efficient MCMC samplers to fit the model chosen by the model selection procedure.  Using a suite of simulations and applied examples, we illustrate how our proposed method outperforms  existing Bayesian methods in terms of both computational efficiency and model selection performance.

The paper is organized as follows. Section~\ref{sec:mp} introduces our formulation of the APLM, including the nonparametric models that we use for the predictor functions and the latent variables that we use to control model selection and smoothness of the predictor functions. For model selection and estimation, appropriate prior distributions are devised in Section~\ref{sec:ps} and computational methods for posterior inference in Section~\ref{sec:pi}. Section~\ref{sec:numill} validates our overall proposed \name method through a suite of numerical examples and simulation studies. Section~\ref{sec:rd} illustrates its application to nutritional epidemiology data and Section~\ref{sec:dc} concludes with discussion. The R package for \name  is currently available at the first author's github and described in Appendix~\ref{app:1}. Proofs of theoretical results are collected in Appendices~\ref{app:2} and~\ref{app:3}.

\section{Formulation of additive partial linear models}
\label{sec:mp}

Under certain smoothness assumptions, the nonlinear additive components, $f_j$,  in~(\ref{eqn:aplm}) can be well approximated by regression splines defined as linear combinations of certain spline basis functions, although the appropriate choice of basis functions depends on the smoothness assumptions. Let ${\mathfrak D}=\{ (y_i,\x_i,\z_i) \}_{i=1}^n$ denote $n$ i.i.d. realizations of $(Y,{\cal X},{\cal Z})$ and let
\begin{align}
	{\cal H}_j^{(n)}=\left\{ h_j \,:\, h_j (\cdot)=\bb_j(\cdot)^\T \bphi_j, ~ \bphi_j\in\mathbb{R}^{M_j},~\sum_{i=1}^n h_j (x_{ij})=0\right\}
	\label{eqn:spline}
\end{align}
denote a spline space spanned by ${M_j}$-dimensional basis functions $\bb_j:\mathbb R\mapsto \mathbb R^{M_j}$ such that $\sum_{i=1}^n h_j (x_{ij})=0$, where $x_{ij}$ is element $j$ of $\x_i$.
We consider a spline approximation $f_j^\ast\in {\cal H}_j^{(n)}$ to the true function $f_j$.
The restriction $\sum_{i=1}^n h_j (x_{ij})=0$ is imposed on the spline space ${\cal H}_j^{(n)}$ in order to replace the identifiability restriction ${\rm E}[f_j(X_j)]=0$ in an empirical sense, i.e., $\sum_{i=1}^n f_j^\ast(x_{ij})=0$. This can be achieved by using the centered basis functions defined by $\bb_j(u)=\bb_j^\ast(u)-n^{-1}\sum_{i=1}^n\bb_j^\ast(x_{ij})$, where $\bb_j^\ast(u)$ denotes a set of unrestricted basis functions.

Many types of basis functions, including B-splines, Bernstein polynomials, Fourier series expansion, and wavelets, are available to estimate a nonlinear function represented by a regression spline, depending on smoothness assumptions. For fast and numerically stable computation, we use the cubic radial basis functions \citep{kohn:etal:01},
\begin{align}
	\bb_j^\ast(u)
	=\left(u,\:|{u-t_{j1}}|^3,\dots,\:|{u-t_{jL_j}}|^3\right)^\T,
	\label{eqn:basis}
\end{align}
for a given $L_j>0$. The abscissae, $\min_{1\le i\le n} x_{ij}< t_{j1}<\dots<t_{jL_j}<\max_{1\le i\le n} x_{ij}$, are candidates for knot locations, only some of which are chosen to achieve acceptable smoothness and curvature of $f_j^\ast$. This construction of basis functions implies that we have $M_j=L_j+1$ for the spline space ${\cal H}_j^{(n)}$ in \eqref{eqn:spline}. Since only a few knots are used among the $L_j$ candidates, the results are not sensitive to a value of $L_j$ unless it is too small. 
It is sufficient that $L_j$ be large enough to ensure that local features
of the function can be detected. We recommend using a value between 20 and 30 for $L_j$ unless the function is known to be extremely wiggly.

We introduce two sets of latent variables into the APLM, one set for controlling the smoothness of nonlinear predictor functions and the other for model selection. First, we let $\bdel_j=(\delta_{jm})_{m=0}^{L_j}$ denote a set of latent variables for smoothness control, where $\delta_{jm}=1$ if basis function $(m+1)$ of the ordered list in~(\ref{eqn:basis}) is used to approximate the nonlinear predictor function $f_j$ and is 0 otherwise. Throughout the paper we use the cardinality notation $|\cdot|$ to denote the number of nonzero elements of a nonnegative vector.
Thus, $|\bdel_{j}|$ denotes the number of basis functions used for $f_j^\ast$, i.e., $|\bdel_{j}|=\sum_{m=0}^{L_j}\delta_{jm}$. Let
$\bb_{\bdel_j}(\cdot)$ denote the $|\bdel_{j}|\times1$ vector of basis functions selected by $\bdel_j$ and let $\bphi_{\bdel_j}$ denote the coefficients corresponding to the basis functions in  $\bb_{\bdel_j}(\cdot)$. Given the latent variables $\bdel_j$, $f_j$ is approximated by the nonlinear function $f_j^\ast$,
\begin{align}
	f_j^\ast(\cdot) = \bb_{\bdel_j}(\cdot)^\T\bphi_{\bdel_j}.
	\label{eqn:fun}
\end{align}
Because there exists $\bphi_j$ such that $\bb_{\bdel_j}(\cdot)^\T\bphi_{\bdel_j}=\bb_{j}(\cdot)^\T\bphi_{j}$ and $\sum_{i=1}^n f_j^\ast (x_{ij})=0$ by the definition of the centered basis function $\bb_j$,  we know that $f_j^\ast \in {\cal H}_j^{(n)}$. 

Turning to the set of latent variables introduced to facilitate model selection in the APLM, let $\bgam^{\rm x}=(\gamma_j^{\rm x})_{j=1}^p$ and $\bgam^{\rm z}=(\gamma_k^{\rm z})_{k=1}^q$ be defined by
\begin{align*}
	\gamma_j^{\rm x}&~=~\left\{
	\begin{array}{l l}
		0 & \text{if $X_j$  has no effect on a response variable}, \\
		1 & \text{if $X_j$ has a linear effect on a response variable},\\
		2 & \text{if $X_j$ has a nonlinear effect on a response variable}, 
	\end{array}\right.\\
	\gamma_k^{\rm z}&~=~\left\{
	\begin{array}{l l}
		0 & \text{if $Z_k$  has no effect on a response variable},\\
		1 & \text{if $Z_k$ has a linear effect on a response variable},
	\end{array}\right.
\end{align*}
for $j=1,\dots,p$ and $k=1,\dots,q$.
The construction of the ternary latent variable $\gamma_j^{\rm x}$ is similar to that of \citet{ross:rubi:19},  but we allow more flexibility by also considering $\gamma_k^{\rm z}$ for the linear components. 
Introducing ternary latent variables makes it easy to assign prior information to each of no effect, linear effect, and nonlinear effect. 
Whereas the additive structure of $\cal Z$ is directly implied by $\bgam^{\rm z}$, determining that of $\cal X$ requires both $\bgam^{\rm x}$ and $\bdel=(\bdel_j)_{j=1}^p$. That is,
all of the basis functions selected by $\bdel_j$ are included in the model if $\gamma_j^{\rm x}=2$, only the linear basis function is included if $\gamma_j^{\rm x}=1$, and none of the basis functions are included if $\gamma_j^{\rm x}=0$.
Thus, we can define a set of new latent variables that indicates whether or not each basis function for $X_j$ is included in a model. 
Specifically,  let
\begin{align}
	\eta(\gamma_j^{\rm x},\bdel_j)=\Big(\mathbbm{1}(\gamma_j^{\rm x}=1)+\delta_{j0}\mathbbm{1}(\gamma_j^{\rm x}=2),~\delta_{j1}\mathbbm{1}(\gamma_j^{\rm x}=2),\dots,~\delta_{jL_j}\mathbbm{1}(\gamma_j^{\rm x}=2)\Big),
	\label{eqn:nla}
\end{align}
where $\mathbbm{1}(\cdot)$ is the indicator function.
That is, element $m$ of $\eta(\gamma_j^{\rm x},\bdel_j)$ is equal to 1 if basis function $m$ in  $\bb_{j}(\cdot)$ is included in a model and is 0 otherwise, so that $\eta(\gamma_j^{\rm x},\bdel_j)$ corresponds to a standard inclusion indicator vector for model selection.
To ensure identifiability, we assume that  if $\gamma_j^{\rm x}=2$, at least one $\delta_{jm}=1$ with $m\ge1$. 

Using the latent variables $\bgam=(\bgam^{\rm x},\bgam^{\rm z})$ and $\bdel$, the APLM with data ${\mathfrak D}$ can be written in matrix form as
\begin{align}
	\y=\alpha\mathbf{1}_n + \B_{\bet(\bgam^{\rm x},\bdel)} \bphi_{\bet(\bgam^{\rm x},\bdel)}+\Z_{\bgam^{\rm z}}\bbe_{\bgam^{\rm z}}+\beps,
	\label{eqn:ammat}
\end{align}
where $\y=(y_1,\cdots,y_n)^\T$ is a vector of response variables, $\B_{\bet(\bgam^{\rm x},\bdel)}$ is the submatrix of the matrix
\begin{align*}
	\B=
	\begin{pmatrix}
		\bb_{1}(x_{11})^\T &  \cdots & \bb_{p}(x_{1p})^\T \\
		\vdots  &  \ddots & \vdots  \\
		\bb_{1}(x_{n1})^\T  & \cdots & \bb_{p}(x_{np})^\T
	\end{pmatrix},
\end{align*}
with columns selected according to $\bet(\bgam^{\rm x},\bdel)=(\eta(\gamma_j^{\rm x},\bdel_j))_{j=1}^p$,
$\bphi_{\bet(\bgam^{\rm x},\bdel)}$ is the vector of coefficients corresponding to the columns of $\B_{\bet(\bgam^{\rm x},\bdel)}$, $\Z_{\bgam^{\rm z}}$ is the submatrix of the matrix $\Z=(\z_1,\dots,\z_n)^\T$ with columns selected according to  $\bgam^{\rm z}$, $\bbe_{\bgam^{\rm z}}$ is a vector of coefficients corresponding to the columns of $\Z_{\bgam^{\rm z}}$, and $\beps \sim N(\mathbf{0}_n,\sigma^2 \mathbf{I}_n)$.

From a practical perspective, how the available predictors are divided into $\mathcal X$ and $\mathcal Z$ is of great interest. We recommend including all continuous variable in 
$\mathcal X$ and including the dummy variables corresponding to all of the
categorical variables as $\mathcal Z$. If there is a strong belief that some continuous variables do not have  nonlinear effects, they can also be treated as a part of $\mathcal Z$.

\section{Prior specification}
\label{sec:ps}

Our ultimate goal is to fit model~(\ref{eqn:ammat}) in a Bayesian framework. Doing this in a computationally efficient manner while maintaining desirable statistical properties requires carefully devised prior distributions. To facilitate formulation of these prior distributions, we let $\W_{[\bgam,\bdel]}=(\B_{\bet(\bgam^{\rm x},\bdel)},{\Z}_{\bgam^{\rm z}})$ and $\bthe_{[\bgam,\bdel]}=(\bphi_{\bet(\bgam^{\rm x},\bdel)}^\T,\bbe_{\bgam^{\rm z}}^\T)^\T$ be a design matrix and the corresponding regression coefficients, respectively, and
let $J_{[\bgam,\bdel]}=|\bet(\bgam^{\rm x},\bdel)|+|\bgam^{\rm z}|$ (defined by $\sum_{j=1}^p \{\mathbbm{1}(\gamma_j^{\rm x}=1)+|\bdel_{j}|\mathbbm{1}(\gamma_j^{\rm x}=2)\}+\sum_{k=1}^q\gamma_k^{\rm z}$) be the dimension of $\bthe_{[\bgam,\bdel]}$.
Without loss of generality, we assume that each column of $\Z$ is centered, so that $\W_{[\bgam,\bdel]}^\T \mathbf{1}_n=\mathbf{0}$; recall that the columns of $\B$ are also centered by definition.

Given the latent variables, $\bgam$ and $\bdel$, model~(\ref{eqn:ammat}) takes the form of a Gaussian linear regression. Thus, the standard conjugate prior distributions for a Gaussian linear regression are {\it semi}-conjugate priors for our APLM since the latent variables are unknown. For model selection, the posterior distribution of $(\bgam,\bdel)$ is of primary interest. Using semi-conjugate prior distributions for the regression parameters allows us to compute the marginal likelihood function given $(\bgam,\bdel)$ analytically -- thus facilitating model selection.  

\subsection{Prior distributions for $\alpha$, $\sigma^2$, and $\bthe_{[\bgam,\bdel]}$}

\label{sec:pmp}
We consider the intercept, $\alpha$, separately from the other regression parameters because it is common to all models. Using an argument based on invariance to scale and location transformations for orthogonal designs \citep{berg:etal:98}, we suggest the improper prior distribution for $(\alpha,\sigma^2)$ given by 
\begin{align*}
	\pi(\alpha,\sigma^2)\propto1/\sigma^2.
\end{align*}
The use of this improper prior is justified since $\alpha$ and $\sigma^2$ appear in all the models we consider,
whereas a proper prior must
be used for $\bthe_{[\bgam,\bdel]}$ \citep{berg:etal:98,baya:etal:12}.
Following convention, we use Zellner's $g$-prior distribution for $\bthe_{[\bgam,\bdel]}$ \citep{zell:86,geor:fost:00,fern:etal:01}. We make the mild assumption
that there is no explicit functional relationship among the columns of $(\X,\Z)$ such that $\W_{[\bgam,\bdel]}$ is full column rank for every $(\bgam,\bdel)$ with $J_{[\bgam,\bdel]}\le n-2$. 
Then for all $J_{[\bgam,\bdel]}\le n-2$, the $g$-prior is given by the normal distribution,
\begin{align*}
	\bthe_{[\bgam,\bdel]}|\W, \bgam,\bdel,g,\sigma^2\sim N\big(0,g\sigma^2(\W_{[\bgam,\bdel]}^\T\W_{[\bgam,\bdel]})^{-1}\big),
\end{align*}
where $g$ is a dispersion factor and  $\W=(\B,\Z)$.
Although the $g$-prior is also defined for $J_{[\bgam,\bdel]}=n-1$ (note that $\W_{[\bgam,\bdel]}^\T\W_{[\bgam,\bdel]}$ is invertible only when $J_{[\bgam,\bdel]}\le n-1$ since the columns of $\W$ are centered), we exclude this case to simplify calculation of 
the marginal likelihood (see \eqref{eqn:ml} and the last paragraph of this subsection below).
Our specification of the joint prior distribution on $(\alpha, \bthe, \sigma^2)$ yields a convenient form for the marginal likelihood ${\cal L}(\y|\W,\bgam,\bdel,g)$.

While the $g$-prior guarantees a closed-form expression for the marginal likelihood, a fixed choice of $g$ would be difficult to determine and
may lead to issues such as Bartlet's paradox or the information paradox \citep{lian:etal:08}. 
Hence we assign a prior to $g$, specifically  
a beta-prime prior distribution \citep{maru:geor:11} with density given by
\begin{align}
	\pi(g;a,b)= \frac{g^b(1+g)^{-a-b-2}}{{B}(a+1,b+1)},\quad g>0,
	\label{eqn:bp}
\end{align}
for $a>-1$ and $b>-1$, where ${B}(\cdot,\cdot)$ is the beta function. Under this prior, $1/(1+g)$ has a beta prior distribution, $Beta(a+1,b+1)$.
As suggested by \citet{maru:geor:11}, we set $a=-3/4$ and $b=(n-J_{[\bgam,\bdel]})/2-7/4$. This choice provides an easy-to-evaluate closed-form expression of the marginalized likelihood (see equation (3.10) and Remark 3.1 of \citet{maru:geor:11}); for every $\bgam$ and $\bdel$ such that $J_{[\bgam,\bdel]}\le n-2$,
\begin{align}
	\begin{split}
		{\cal L}(\y|\W,\bgam,\bdel)&=\int {\cal L}(\y|\W,\bgam,\bdel,g)\pi(g;-3/4,(n-J_{[\bgam,\bdel]})/2-7/4)d g\\
		&=\frac{K(n,\y)\,{B}(J_{[\bgam,\bdel]}/2+1/4,(n-J_{[\bgam,\bdel]})/2-3/4)}{\big(1-R_{[\bgam,\bdel]}^2\big)^{(n-J_{[\bgam,\bdel]})/2-3/4}\,{B}(1/4,(n-J_{[\bgam,\bdel]})/2-3/4)},
	\end{split}
	\label{eqn:ml}
\end{align}
where $R_{[\bgam,\bdel]}^2$ is the coefficient of determination of the regression model determined by $(\bgam,\bdel)$, and $K(n,\y)={\pi}^{-(n-1)/2}\,n^{1/2}\, \Gamma ((n-1)/2)\, \lVert\y-\bar y\mathbf{1}_n\rVert^{-(n-1)}$ with $\bar y=n^{-1}\sum_{i=1}^n y_i$.

The prior in \eqref{eqn:bp} is not proper if $J_{[\bgam,\bdel]}=n-1$ with our choice of $a$ and $b$. Although it is still possible to calculate a
marginal likelihood by using 
a proper prior in this particular case (equation (3.9) of \citet{maru:geor:11}), we instead exclude the case $J_{[\bgam,\bdel]}=n-1$ for simplicity. Hence, we impose the restriction $J_{[\bgam,\bdel]}\le n-2$ in the construction of the priors for $\bgam$ and $\bdel$ throughout the following subsections. We close this subsection by noting that the posterior distribution is proper despite the use of the improper prior $\pi(\alpha,\sigma^2)\propto1/\sigma^2$. This can be easily checked by inspection of the expression for the marginal likelihood in~\eqref{eqn:ml}.

\subsection{Prior distribution for $\bgam$}
\label{sec:pga}
In this and the next subsections, we specify the prior distribution on the latent variables $\bgam$ and $\bdel$. Throughout we impose the restriction $J_{[\bgam,\bdel]}\le n-2$ as mentioned above, i.e., we assume $\pi(J_{[\bgam,\bdel]}\le n-2)=1$.

A popular prior distribution for latent indicator variables in Bayesian variable selection is the so-called hierarchical uniform prior, a mixture of a uniform distribution and a product of Bernoulli distributions \citep{kohn:etal:01,crip:etal:05}. The probability mass function for 
a binary vector $\v=(v_1,\dots,v_r)^\T$ following the hierarchical uniform distribution is given by $\pi(\v)={{B}\left(\sum_{i=1}^r v_i+1,r-\sum_{i=1}^r v_i+1\right)}$.
The distribution is considered to be vaguely informative because it implies $\sum_{k=1}^r v_k$ is uniformly distributed on $0, 1, \ldots, r$. In variable selection, the distribution implies a discrete uniform distribution on the number of variables, provides effective multiplicity adjustment \citep{scot:berg:10}, and outperforms other prior distributions such as a product of Bernoulli distributions with fixed probability \citep{more:etal:15,wang:maru:17}.

In our setting, however, the elements of $\bgam^{\rm x}$ are ternary (rather than binary).  To accommodate this while preserving the desirable properties of the hierarchical uniform prior distribution, we propose a prior distribution for $\bgam$, given by
\begin{align}
	\pi(\bgam)&=\frac{1}{2^{|\bgam^{\rm x}|}} B \left(|\bgam|+1,~p+q-|\bgam|+1\right),\quad \bgam:0\le |\bgam|\le p+q,
	\label{eqn:bb}
\end{align}
where $|\bgam|=\sum_{j=1}^p\mathbbm{1}(\gamma_j^{\rm x}>0)+\sum_{k=1}^q\gamma_k^{\rm z}$ and $|\bgam^{\rm x}|=\sum_{j=1}^p\mathbbm{1}(\gamma_j^{\rm x}>0)$ (following the definition of $|\cdot|$ given in Section~\ref{sec:mp}). As with the hierarchical uniform prior distribution, further insight can be obtained by considering the prior distribution of $|\bgam|$ induced by~(\ref{eqn:bb}). This is explored in the following proposition. 
\begin{prop}
	The prior distribution in (\ref{eqn:bb}) implies $\pi(|\bgam|=k)=1/(p+q+1)$ for $k=0,1,\dots,p+q$.
	\label{thm:pr1}
\end{prop}
\noindent 
Proposition~\ref{thm:pr1}  states that the prior distribution in~(\ref{eqn:bb}) assigns equal probabilities to the number of predictors in the model in the same way that the hierarchical uniform prior distribution does.  Thus, the prior in (\ref{eqn:bb}) is considered to be weakly informative in the same sense that the hierarchical uniform prior distribution is. The use of the prior distribution in~(\ref{eqn:bb}) can also be supported by examining its marginal prior distributions for $\gamma_j^{\rm x}$ and $\gamma_k^{\rm z}$.
\begin{prop}
	The marginal prior distributions of $\gamma_j^{\rm x}$ and $\gamma_k^{\rm z}$ induced by the prior distribution in (\ref{eqn:bb}) satisfy $\pi(\gamma_j^{\rm x}=0)=1/2$, $\pi(\gamma_j^{\rm x}=1)=\pi(\gamma_j^{\rm x}=2)=1/4$ for $j=1,\dots,p$, and $\pi(\gamma_k^{\rm z}=0)=\pi(\gamma_k^{\rm z}=1)=1/2$ for $k=1,\dots,q$, respectively.
	\label{thm:pr2}
\end{prop}
\noindent 
Proposition~\ref{thm:pr2} states that the prior distribution in~(\ref{eqn:bb}) implies that all of the variables are equally likely to be excluded or included in the model, and when they are included, the potentially nonlinear variables are equally likely to have a linear or nonlinear effect. Proofs for Propositions~\ref{thm:pr1} and~\ref{thm:pr2} are provided in Appendices~\ref{app:2} and~\ref{app:3}, respectively.

\subsection{Prior distribution for $\bdel$}
\label{sec:pde}

It is sensible to assume independent prior distributions for different  predictor functions, and we do so conditional on the order of the predictor function (no, linear, or nonlinear effect) as recorded by $\bgam^{\rm x}$. Specifically, we assume
$\pi(\bdel|\bgam^{\rm x})=\prod_{j=1}^p \pi(\bdel_j|\gamma_j^{\rm x})$. 
Partitioning 
$\bdel_j$ into $(\delta_{j0},\bdel_{j\backslash0})$  with $\bdel_{j\backslash0}=(\delta_{jm})_{m=1}^{L_j}$,
we assume that $\delta_{j0}$ and $\bdel_{j\backslash0}$ are a priori independent, i.e., $\pi(\bdel_j|\gamma_j^{\rm x})=\pi(\delta_{j0}|\gamma_j^{\rm x})\,\pi(\bdel_{j\backslash0}|\gamma_j^{\rm x})$. 

Because the basis functions for predictor function $j$ only come into the model  when $\gamma_j^{\rm x} =2$, the latent variables $\bdel_j$ do not affect the model if $\gamma_j^{\rm x}\ne2$.
Still we specify both of the prior distributions $\pi(\bdel_j|\gamma_j^{\rm x}\ne2)$ and $\pi(\bdel_j|\gamma_j^{\rm x}=2)$ to enable efficient sampling of the posterior distribution; see Section~\ref{sec:332}. Of course, regardless of the observed data ${\mathfrak D}$,  we have $\pi(\bdel_j|\gamma_j^{\rm x}\ne2) = \pi(\bdel_j|\gamma_j^{\rm x}\ne2, {\mathfrak D})$. This type of a prior distribution is often called a ``pseudo-prior distribution'' \citep{carl:chib:95,dell:etal:02}. Unless otherwise specified, when we refer to the prior distribution of $\bdel_j$, we mean $\pi(\bdel_j|\gamma_j^{\rm x}=2)$. 

\subsubsection{Prior distributions for nonlinear predictor functions, $\pi(\bdel_j|\gamma_j^{\rm x}=2)$}
\label{sec:331}
To begin with, we specify that a priori probability of each linear basis function being included for nonlinear predictor functions is equal to 50\%, i.e., we simply set $\pi(\delta_{j0}|\gamma_j^{\rm x}=2)=1/2$ for each $j$.  The situation becomes more complicated for nonlinear basis functions. Because $\gamma_j^{\rm x}=2$ indicates that predictor function $j$ is nonlinear, its decomposition must include at least one nonlinear basis function; otherwise the model is not identifiable in view of \eqref{eqn:nla}. This demands $\pi(|\bdel_{j\backslash0}| > 0 \,|\, \gamma_j^{\rm x}=2) = 1$, where $|\bdel_{j\backslash0}| = \sum_{m=1}^{L_j}\delta_{jm}$. 
We achieved this via a zero-truncated prior distribution formed by restricting a distribution for $\bdel_{j\backslash0}$ to the set $\{\bdel_{j\backslash0}:1\le |\bdel_{j\backslash0}|\le L_j\}$.

The theory of random series prior distributions suggests that a prior on $|\bdel_{j\backslash0}|$ should decay to zero to ensure optimal posterior contraction \citep{beli:serr:14,shen:ghos:15}.
Although we use a different basis, with this in mind, we set a geometric prior distribution on $|\bdel_{j\backslash0}|$, renormalized so that $1\le|\bdel_{j\backslash0}|\le L_j$, and assume that all values of $\bdel_{j\backslash0}$ resulting in the same value of  $|\bdel_{j\backslash0}|$ are equally likely. 
That is, we assume 
\begin{align}
	\pi(\bdel_{j\backslash 0}|\gamma_j^{\rm x}=2)={{L_j}\choose{|\bdel_{j\backslash0}|}}^{-1}\frac{(1-p_j)^{|\bdel_{j\backslash0}|-1}p_j}{1-(1-p_j)^{L_j}},\quad  \bdel_{j\backslash0}: 1\le |\bdel_{j\backslash0}|\le L_j,
	\label{eqn:pde}
\end{align}
where $0<p_j<1.$
The prior distribution of $|\bdel_{j\backslash0}|$ induced by~(\ref{eqn:pde}) approaches a zero-truncated geometric distribution as $L_j$ increases.
Based on our experience, we recommend using a fixed value of $p_j$ between 0.3 and 0.6 so that the prior on $|\bdel_{j\backslash0}|$ 
is sufficiently concentrated on values less than 10 to 15. We use a fixed choice $p_j=0.4$ in all of the numerical analyses in this article.

\subsubsection{Specifying the pseudo-prior distributions $\pi(\bdel_j|\gamma_j^{\rm x}\ne2)$}
\label{sec:332}

Recall that from~(\ref{eqn:nla}), the marginal likelihood is invariant to $\bdel_j$ if $\gamma_j^{\rm x}\ne2$ and thus the pseudo-prior distribution, $\pi(\bdel_j|\gamma_j^{\rm x}\ne2)$, is irrelevant to the marginal posterior distribution of $\bgam$, i.e.,
\begin{align*}
	\pi(\bgam|{\mathfrak D})&\propto\pi(\bgam)\sum_{\bdel} {\cal L}(\y|\W,\bgam,\bdel)\pi(\bdel|\bgam^{\rm x})=\pi(\bgam)\sum_{\bdel_j:\gamma_j^{\rm x}=2}\Bigg\{{\cal L}(\y|\W,\bgam,\bdel)\prod_{j:\gamma_j^{\rm x}=2}\pi(\bdel_j|\gamma_j^{\rm x}=2)\Bigg\}.
\end{align*}
Although the choice of a pseudo-prior distribution does not affect the posterior distribution, it does affect the convergence of our posterior sampling algorithms. In particular, if $\pi(\bdel_{j\backslash0}|\gamma_j^{\rm x}\ne2)$ is similar to $\pi(\bdel_{j\backslash0}|\gamma_j^{\rm x}=2, {\mathfrak D})$, then $\gamma_j^{\rm x}$ jumps more frequently in our MCMC sampler without being stuck at $\gamma_j^{\rm x}=0$ or $\gamma_j^{\rm x}=1$. For this reason, 
it is often recommended that a pilot chain be run to examine the posterior distribution of model-specific parameters 
\citep{dell:etal:02, vand:etal:09}. Following this advice, we examine $\pi(\bdel|\bgam_F,{\mathfrak D})$ by running a pilot chain using \eqref{eqn:ml} and \eqref{eqn:pde}, where $\bgam_{F}$ denotes a full model with $\gamma_j^{\rm x}=2$, $j=1,\dots,p$, and $\gamma_k^{\rm z}=1$, $k=1,\dots,q$. We then set $\pi(\bdel_j|\gamma_j^{\rm x}\ne2) =\hat\pi(\delta_{j0}|\bgam_F,{\mathfrak D})\,\hat\pi(\bdel_{j\backslash0}|\bgam_F,{\mathfrak D})$, where $\hat\pi$ denotes an estimated probability distribution obtained from the pilot chain. The first factor in the approximation can easily be computed as the relative frequency of $\delta_{j0}=0$ and $\delta_{j0}=1$, for each $j$, with the constraint that neither probability may be zero or one. (We truncate $\hat\pi(\delta_{j0}|\bgam_F,{\mathfrak D})$ so
that it is between 0.01 and 0.99.)

Due to the dimensionality of $\bdel_{j\backslash0}$, the pilot chain may not visit all areas of its parameter space, making approximating the joint posterior distribution $\pi(\bdel_{j\backslash0}|\bgam_F,{\mathfrak D})$ challenging. (We must consider the joint distribution because adjacent nonlinear spline basis functions are typically highly correlated.)
To circumvent this issue, we smooth the posterior distribution of $\bdel_{j\backslash0}$ obtained by the pilot chain. Specifically, we use a normal density function to approximate the posterior distribution of $|\bdel_{j\backslash0}|$, and then put equal weight on each value of $\bdel_{j\backslash0}$ that gives the same value of $|\bdel_{j\backslash0}|$. (Note that we smooth the discrete distribution using a continuous density.) This ensures a nonzero density on the entire parameter space. The resulting smoothed posterior distribution is given by
\begin{align}
	\hat\pi(\bdel_{j\backslash0}|\bgam_F,{\mathfrak D})
	={{L_j}\choose{|\bdel_{j\backslash0}|}}^{-1} \frac{\exp\{-(|\bdel_{j\backslash0}|-\xi_j)^2/2\nu_j^2\}}{\sum_{\ell=1}^{L_j}\exp\{-(\ell-\xi_j)^2/2\nu_j^2\}}, \quad
	\bdel_{j\backslash0} : 1\le|\bdel_{j\backslash0}|\le L_j,
	\label{eqn:disn}
\end{align}
where the hyperparameters $\xi_j$ and $\nu_j^2$ are set by solving nonlinear equations to match the posterior mean and variance of $|\bdel_{j\backslash0}|$ obtained by the pilot chain to those of the distribution in~(\ref{eqn:disn}). 
(If the posterior variance of $|\bdel_{j\backslash0}|$ obtained by the pilot chain is zero, we adjust it to be strictly positive in matching the hyperparameters.)
Smoothing in this way allows us to run a shorter pilot chain that may not have fully converged. 
Because $|\bdel_{j\backslash0}|$ is discrete and restricted on $1\le|\bdel_{j\backslash0}|\le L_j$, the hyperparameters  $\xi_j$ and $\nu_j^2$ are not the (approximated) posterior mean and variance of $|\bdel_{j\backslash0}|$, but values that reproduce them using~(\ref{eqn:disn}). Based on our experience, the density in~(\ref{eqn:disn}) gives good empirical results compared with other parametric smoothing alternatives. 

The above construction of the pseudo-prior requires $p+q\le n-2$ due to the use of $\bgam_F$. 
Even if $p+q> n-2$, a pilot chain can still be run with a minor modification. This can be done, for example, by setting $\gamma_k^{\rm z}=0$, $k=1,\dots,q$, for $\bgam_F$ and by ignoring the restriction $|\bdel_{j\backslash0}| > 0$ (there is no identifiability issue due to fixed $\bgam_F$). Our experience shows that this approach results in slightly poorer convergence, but not significantly so.
Our R package deploys this strategy when $p+q> n-2$.

\section{Monte Carlo sampling algorithms}
\label{sec:pi}

In this section we present the three Monte Carlo sampling algorithms needed for \namenospace:  (i) the algorithm used to obtain the pilot chain needed to set the pseudo-prior distribution described in Section~\ref{sec:332}, (ii) the algorithms used to sample the posterior distribution of $\bgam$ for model selection and (iii) the algorithm used to sample the model parameters of the selected model from the posterior distribution. 

\subsection{Obtaining the pilot chain needed for the pseudo-prior distribution}
\label{sec:pseudo}
To specify the pseudo-prior distribution, $\pi(\bdel_j|\gamma_j^{\rm x}\ne2)$, we require a pilot chain that is a realization of a Markov chain with stationary distribution $p(\bdel | \bgam_F, {\mathfrak D})$; see Section~\ref{sec:332}. We obtain the Markov chain via a Metropolis within blocked Gibbs sampler.
In our experience, a small number of iterations (e.g., 100 to 200) is enough for a pilot chain to find a reasonable approximation for the pseudo-prior distribution.

\medskip
\noindent 
{\it Sampler 1: Metropolis within blocked Gibbs sampler for pilot chain}

\smallskip
Given starting values $\bdel^{(0)}$, iterate the following.

\smallskip
For iteration $t = 1, 2, \dots$:
\begin{enumerate}[leftmargin=15mm]
	\item Partition the $(p+\sum_{j=1}^p L_j)$ components of $\bdel$ into blocks. First, separate each of $\delta_{10}, \ldots, \delta_{p0}$ into its own block of size one; in this way the sampler updates the linear and nonlinear basis functions separately. 
	Randomly partition the remaining $L_j$ elements of each $\bdel_{j\backslash0}$ into blocks of size 2, 3, or 4. The procedure partitions $\bdel$ into a set of blocks, $[\bdel]_1,\dots,[\bdel]_B$, where $B$ is the number of blocks constructed in iteration $t$. An example of possible blocks is 
	\begin{align*}
		& (\delta_{10}),  (\delta_{11},\delta_{12},\delta_{13},\delta_{14}),  (\delta_{15},\delta_{16}),  \dots,  (\delta_{1L_1-2},\delta_{1L_1-1},\delta_{1L_1}),  \\
		& \qquad\qquad\qquad\qquad\qquad\qquad\qquad\vdots \\
		& (\delta_{j0}),  (\delta_{j1},\delta_{j2}), \ (\delta_{j3},\delta_{j4}, \delta_{j5}), \dots,  (\delta_{j,L_j-3},\delta_{j,L_j-2},\delta_{j,L_j-1},\delta_{jL_j}), \\
		& \qquad\qquad\qquad\qquad\qquad\qquad\qquad\vdots\\
		& (\delta_{p0}),  (\delta_{p1},\delta_{p2}, \delta_{p3}),  (\delta_{p4}, \delta_{p5},\delta_{p6}),  \dots,  (\delta_{pL_p-2}, \delta_{pL_p-1},\delta_{pL_p}) . 
	\end{align*}
	Assign the current value in $\bdel^{(t-1)}$ to each of the block to arrive at $[\bdel^{(t-1)}]_1, \ldots [\bdel^{(t-1)}]_B$.
	
	\item For block $b=1, \ldots, B$:
	\begin{enumerate}
		\item Sample a proposal $[\bdel^\prime]_b$ of $[\bdel]_b$ from the conditional prior distribution induced by $\pi(\bdel|\bgam_F)$, i.e., sample  
		\begin{align*}
			[\bdel^\prime]_b \sim \pi( [\bdel]_b \,|\,[\bdel^{(t)}]_{-b},\bgam_F)\propto \pi( [\bdel]_b ,[\bdel^{(t)}]_{-b} | \bgam_F),
		\end{align*}
		where 
		$[\bdel^{(t)}]_{-b} = ([\bdel^{(t)}]_1, \ldots, [\bdel^{(t)}]_{b-1}, [\bdel^{(t-1)}]_{b+1}, \ldots, [\bdel^{(t-1)}]_{B})$.
		\item Compute the acceptance probability 
		\begin{align*}
			\label{eqn:acprob}
			r= \min\left\{1,\frac{{\cal L}(\y|\W,\bgam_F,[\bdel^\prime]_b,[\bdel^{(t)}]_{-b})}{{\cal L}(\y|\W,\bgam_F,[\bdel^{(t)}]_b,[\bdel^{(t)}]_{-b})}\right\}.
		\end{align*}
		\item Set 
		$[\bdel^{(t)}]_b =
		\begin{cases}
			[\bdel^\prime]_b & \hbox{with probability } r, \\
			[\bdel^{(t-1)}]_b & \hbox{otherwise}.
		\end{cases}
		$
	\end{enumerate}
\end{enumerate}

Since adjacent nonlinear basis functions are highly correlated, individual updating algorithms such as the standard Gibbs sampler may not work very well. Thus, Sampler~1 uses a blocking scheme to enable the swapping of
adjacent basis functions 
and avoid the chain being trapped
in a local mode.
On the other hand, the linear basis term is not directly associated with other nonlinear basis functions. Since it is not clear how they are correlated, each $\delta_{j0}$ is isolated by forming its own block of size one.
The partition changes every iteration for more flexibility.

In a nonparametric regression with basis selection, a small number of basis functions is often sufficient to estimate an unknown function, and hence the current value of $[\bdel^{(t)}]_b$ is likely to be sparse. We thus use the conditional prior with a decaying tail property as a proposal distribution for $[\bdel]_b$, so that a zero (or sparse) vector is likely to be proposed for each block. This allows us to circumvent the time-consuming evaluation of the acceptance probability $r$ when the proposal is identical to the current value of $[\bdel^{(t)}]_b$. Using the conditional prior as a proposal distribution improves the convergence efficiency of Sampler~1 with respect to {\it per floating point operation} rather than {\it per iteration} \citep{kohn:etal:01}. \citet{kohn:etal:01} empirically presents the efficiency of a sampling scheme based on a similar idea.

\subsection{Bayesian model selection}
\label{sec:bvs}
For model selection we obtain a Monte Carlo sample from the marginal posterior distribution $\pi(\bgam|{\mathfrak D})$.  We do so by using a Metropolis within (blocked) Gibbs sampler to obtain
a Markov chain with stationary distribution equal to $\pi(\bgam,\bdel|{\mathfrak D})$ and discarding the sample of~$\bdel$. (Owing to the pseudo-prior distribution the sample of $\bdel$ does not have an immediate interpretation and is only introduced as a mechanism to obtain the sample of $\bgam$.) For better readability, the values of ${\gamma_j^{\rm x}}$ and ${\gamma_k^{\rm z}}$ at iteration $t$ of the following sampler are written as $[{\gamma_j^{\rm x}}]^{(t)}$ and $[{\gamma_k^{\rm z}}]^{(t)}$, respectively.

\medskip
\noindent 
{\it Sampler 2: Generating a Markov chain with stationary distribution $\pi(\bgam,\bdel|{\mathfrak D})$}

\smallskip
Given starting values $[\bgam^{\rm x}]^{(0)}=([\gamma_j^{\rm x}]^{(0)} )_{j=1}^p$, $[\bgam^{\rm z}]^{(0)}=( [\gamma_k^{\rm z}]^{(0)} )_{k=1}^q$ and $\bdel^{(0)}=(\bdel_j^{(0)})_{j=1}^p$, iterate the following.

\smallskip
For iteration $t = 1, 2, \dots$:

\begin{enumerate}[leftmargin=15mm] 
	\item Partition the index set $\{1,\dots,p\}$ into two disjoint sets, $\mathcal J_1^{(t)}=\{j:[\gamma_j^{\rm x}]^{(t-1)}\ne2\}$ and $\mathcal J_2^{(t)}=\{j:[\gamma_j^{\rm x}]^{(t-1)}=2\}$.
	\item
	Draw $\{\bdel_j^{(t)}:j\in\mathcal J_1^{(t)}\}$ from the constructed pseudo-prior,
	\begin{align*}
		\prod_{j\in\mathcal J_1^{(t)}}\pi(\bdel_j|\gamma_j^{\rm x}\ne2)=\prod_{j\in\mathcal J_1^{(t)}}\hat\pi(\delta_{j0}|\bgam_F,{\mathfrak D})\,\hat\pi(\bdel_{j\backslash0}|\bgam_F,{\mathfrak D}).
	\end{align*}
	\item
	Draw $\{\bdel_j^{(t)}:j\in\mathcal J_2^{(t)}\}$ using Sampler~1, but with the current value $\bgam^{(t-1)}$ instead of fixed $\bgam_F$. (Unlike in Sampler~1, we do not update all the components of $\bdel$ here.)
	\item
	For $j=1,\dots,p$, draw $[\gamma_j^{\rm x}]^{(t)}$ from the conditional posterior $\pi(\gamma_j^{\rm x}|[\bgam_{-j}^{\rm x}]^{(t)},[{\bgam^{\rm z}}]^{(t-1)},\bdel^{(t)},\mathfrak D)$, which is a discrete distribution with probabilities,
	\begin{align*}
		&\pi(\gamma_j^{\rm x}=\ell|\bgam^{\rm x}_{-j},{\bgam^{\rm z}},\bdel,\mathfrak D) \\
		&\quad\propto
		{\cal L}(\y|\W,\gamma_j^{\rm x}=\ell,\bgam^{\rm x}_{-j},{\bgam^{\rm z}},\bdel)\,\pi(\gamma_j^{\rm x}=\ell,\bgam^{\rm x}_{-j},{\bgam^{\rm z}})\,\pi(\bdel_j|\gamma_j^{\rm x}=\ell),\quad\ell=0,1,2,
	\end{align*}
	where $[\bgam_{-j}^{\rm x}]^{(t)}=([{\gamma_1^{\rm x}}]^{(t)},\dots ,[{\gamma_{j-1}^{\rm x}}]^{(t)},[{\gamma_{j+1}^{\rm x}}]^{(t-1)},\dots , [{\gamma_{p}^{\rm x}}]^{(t-1)})$, $\bgam^{\rm x}_{-j}=(\gamma_{j'}^{\rm x})_{j'\ne j}$, and $\pi(\bdel_j|\gamma_j^{\rm x}=\ell)$ is the prior in Section~\ref{sec:331} if $\ell=2$ and the pseudo-prior in Section~\ref{sec:332} otherwise.
	\item For $k=1,\dots,q$, draw $[{\gamma_k^{\rm z}}]^{(t)}$ from the conditional posterior $\pi(\gamma_k^{\rm z}|
	[\bgam^{\rm x}]^{(t)},[{\bgam_{-k}^{\rm z}}]^{(t)},\bdel^{(t)},\mathfrak D)$,
	which is a Bernoulli distribution with probabilities,
	\begin{align*}
		\pi(\gamma_k^{\rm z}=\ell|\bgam^{\rm x},{\bgam^{\rm z}_{-k}},\bdel,\mathfrak D) \propto
		{\cal L}(\y|\W,\bgam^{\rm x},\gamma_k^{\rm z}=\ell,\bgam_{-k}^{\rm z},\bdel)\,\pi(\bgam^{\rm x},\gamma_k^{\rm z}=\ell,\bgam_{-k}^{\rm z}),\quad\ell=0,1,
	\end{align*}
	where $[\bgam_{-k}^{\rm z}]^{(t)}=([{\gamma_1^{\rm z}}]^{(t)},\dots, [{\gamma_{k-1}^{\rm z}}]^{(t)},[{\gamma_{k+1}^{\rm z}}]^{(t-1)}, \dots , [{\gamma_{p}^{\rm z}}]^{(t-1)})$ and $\bgam^{\rm z}_{-k}=(\gamma_{k'}^{\rm z})_{k'\ne k}$.
\end{enumerate}

An estimate of $\bgam$, denoted by $\hat\bgam$, can be obtained with the accumulated MCMC samples. 
Because some values of $\bgam$ are ternary, ordinary Bayesian model selection criteria such as the median probability model \citep{barb:etal:04} are 
not directly applicable. To find the most likely model, we thus generalize the median probability model such that $\hat\bgam$ is determined by the marginal 
maximum a posteriori
estimates: $\hat\gamma_j^{\rm x}=\operatorname*{arg\,max}_{\ell\in\{0,1,2\}} \pi(\gamma_j^{\rm x}=\ell|{\mathfrak D})$, $j=1\dots,p$, and $\hat\gamma_k^{\rm z}=\operatorname*{arg\,max}_{\ell\in\{0,1\}} \pi(\gamma_k^{\rm z}=\ell|{\mathfrak D})$, $k=1\dots,q$, where the marginal posteriors are estimated by the collected MCMC draws.

\subsection{Estimation after model selection}
\label{sec:est}
Once we have found the most likely model, as specified by $\hat\bgam$, we can fit this model by obtaining a Markov chain with stationary distribution $\pi(\alpha,\bthe_{[\hat\bgam,\bdel]},\sigma^2,g, \bdel | \hat\bgam, {\mathfrak D})$. 

Here we present an efficient sampler for APLMs using the data-driven knot-selection technique described in Section~\ref{sec:intro}. 
The convergence characteristics of the sampling scheme can be improved by collapsing the parameters $(\alpha,\bthe_{[\hat\bgam,\bdel]},\sigma^2,g)$ out of the model. In the algorithm below, we update $\bdel_j$ only for $j$ such that $\hat\gamma_j^{\rm x} = 2$ because the latent variables $\bdel_j$ affect the model likelihood only for such $j$. We still use the full vector notation $\bdel$ for simplicity, but the actual target distribution $\pi(\alpha,\bthe_{[\hat\bgam,\bdel]},\sigma^2,g, \bdel | \hat\bgam, {\mathfrak D})$ is obtained by drawing $\bdel_j$ for $j$ such that $\hat\gamma_j^{\rm x} \ne 2$ together using the pseudo-prior distribution.

\medskip
\noindent 
{\it Sampler 3: Generating a Markov chain with stationary distribution $\pi(\alpha,\bthe_{[\hat\bgam,\bdel]},\sigma^2,g ,\bdel|\hat\bgam, {\mathfrak D})$}

\smallskip
Given starting values $\bdel^{(0)}=(\bdel_j^{(0)})_{j=1}^p$, iterate the following.

\smallskip
For iteration $t = 1, 2, \dots$:

\begin{enumerate}[leftmargin=15mm]
	
	\item For $\hat{\mathcal J}_2=\{j:\hat\gamma_j^{\rm x}=2\}$, draw $\{\bdel_j^{(t)}:j\in\hat{\mathcal J}_2\}$ using Sampler~1, but with the estimate $\hat\bgam$ instead of $\bgam_F$.
	
	\item
	Draw $g_\ast^{(t)}$ from $\pi(g_\ast|\hat\bgam,\bdel^{(t)},{\mathfrak D})$ which is a beta distribution,
	\begin{align*}
		g_\ast|\hat\bgam,\bdel,{\mathfrak D}~\sim~{Beta}\left(\frac{J_{[\hat\bgam,\bdel]}}{2}+\frac{1}{4},\frac{n-J_{[\hat\bgam,\bdel]}}{2}-\frac{3}{4}\right),
	\end{align*}
	and let $g^{(t)}=(1/g_{\ast}^{(t)}-1)\big/\big(1-R_{[\hat\bgam,\bdel^{(t)}]}^2\big)$.
	\item
	Draw $[{\sigma^2}]^{(t)}$ from $\pi(\sigma^2|\hat\bgam,\bdel^{(t)},g^{(t)},{\mathfrak D})$ which is an inverse gamma distribution,
	\begin{align*}
		\sigma^2|\hat\bgam,\bdel,g,{\mathfrak D}~\sim~{ IG}\left(\frac{n-1}{2},~\frac{\lVert\y-\bar y\mathbf{1}_n\rVert^2\big[1+g\big(1-R_{[\hat\bgam,\bdel]}^2\big)\big]}{2(1+g)}\right).
	\end{align*}
	\item
	Draw $(\alpha^{(t)},\bthe_{[\hat\bgam,\bdel^{(t)}]}^{(t)})$ from $\pi(\alpha,\bthe_{[\hat\bgam,\bdel^{(t)}]}|\hat\bgam,\bdel^{(t)},[{\sigma^2}]^{(t)},g^{(t)},{\mathfrak D})$ which is a product of two independent normal distributions,
	\begin{align*}
		\alpha|\hat\bgam,\bdel,\sigma^2,g,{\mathfrak D}&~\sim~{ N}\left(\bar y , n^{-1}\sigma^2\right),\\
		\bthe_{[\hat\bgam,\bdel]}|\hat\bgam,\bdel,\sigma^2,g,{\mathfrak D}&~\sim~ {N} \left(\frac{g}{1+g}\tilde\bthe_{[\hat\bgam,\bdel]},~\frac{g \sigma^2}{1+g} \left(\W_{[\hat\bgam,\bdel]}^\T \W_{[\hat\bgam,\bdel]}\right) ^{-1}\right),
	\end{align*}
	where $\tilde\bthe_{[\hat\bgam,\bdel]}=\big(\W_{[\hat\bgam,\bdel]}^\T  \W_{[\hat\bgam,\bdel]}\big)^{-1} \W_{[\hat\bgam,\bdel]}^\T  (\y-\bar y\mathbf{1}_n)$.
\end{enumerate}

The conditional distributions sampled by Sampler~3 are functionally incompatible  \citep[as defined in Section~2.2 of][]{hobe:case:98}. This means that there is no joint distribution for which the sampled conditionals are the full conditional distribution and changing the order of the draws may alter the stationary distribution of the resulting Markov chain 
\citep{vand:park:08, park:lee:21}. Formally, the sampled distributions are conditional distributions of a nested sequence of marginal distributions of the target posterior, i.e., a sequence of partially collapsed distributions \citep{vand:park:08}. In this case, it is straightforward to verify that the sampler's stationary distribution is indeed $\pi(\alpha,\bthe_{[\hat\bgam,\bdel]},\sigma^2,g ,\bdel|\hat\bgam, {\mathfrak D})$. Step~1 (combined with sampling all other entries of $\bdel$ using the pseudo-prior) samples a set of complete conditional distributions of the marginal posterior distribution $\pi(\bdel|\hat\bgam, {\mathfrak D})$ and operates completely independently of Steps~2--4.  Thus, its (Metropolis within blocked Gibbs) sampler maintains the marginal target distribution, $\pi(\bdel|\hat\bgam, {\mathfrak D})$, as its stationary distribution \citep[e.g.,][]{vand:jiao:15}. The subsequent steps directly sample the conditional distributions 
$\pi(g  | \bdel, \hat\bgam, {\mathfrak D})$,
$\pi(\sigma^2 | g ,\bdel, \hat\bgam, {\mathfrak D})$, and 
$\pi(\alpha,\bthe_{[\hat\bgam,\bdel]} | \sigma^2,g ,\bdel, \hat\bgam, {\mathfrak D})$, each of which maintains the target posterior distribution. 
In general, care must be taken when implementing algorithms such as Sampler~3 that involve partially collapsed conditional distributions to be sure that the resulting stationary distribution coincides with the target (posterior) distribution; see \citet{park:vand:09} for details.

Sampler 3 enables us to fit the APLM with a given set of covariates while averaging over the knot indicator variables, $\bdel$, which change over the Markov chain iterations \citep[i.e., Bayesian model averaging,][]{raft:etal:97}.  More specifically, model-averaged estimates of smooth functions are obtained by recovering $f_j^\ast$ using~(\ref{eqn:fun}) with posterior samples of $\bdel_j$ and $\bphi_{\bdel_j}$ in each iteration; that is, with $T$ iterations of a Markov chain, we obtain
$$\hat f_j^\ast(\cdot) = \frac{1}{T}\sum_{t=1}^T\bb_{\bdel_j^{(t)}}(\cdot)^\T\bphi_{\bdel_j^{(t)}}^{(t)}$$ 
for $j$ such that $\hat\gamma_j^{\rm x}=2$. This estimate reduces the bias caused by fixing a single set of basis functions. We could also average over $\bgam$ rather than conditioning on the estimate $\hat\bgam$. This approach is known to work well for prediction \citep{raft:etal:97}, but it is more difficult to interpret the results since this approach does not produce a set of included covariates.  We do not consider this form of model averaged prediction in this article.

As a final remark, we recall that the procedure requires $\Z$ to be centered for identifiability. This can be done by the translation $\Z=\Z^
\ast-n^{-1}\mathbf{1}_n\mathbf{1}_n^\top \Z^\ast$ with the original (uncentered) matrix $\Z^\ast$.
Since this operation changes the intercept in the model, the intercept of the original location, denoted by $\alpha^
\ast$, can be restored by the translation $\alpha^\ast=\alpha-\mathbf{1}_n^\top{\Z}_{\hat\bgam^{\rm z}}^\ast \bbe_{\hat\bgam^{\rm z}}$ using the uncentered $\Z^\ast$ and samples of $\alpha$ and $\bbe_{\hat\bgam^{\rm z}}$ in each MCMC iteration.

\section{Numerical illustration}
\label{sec:numill}

We conduct a simulation study to evaluate the performance of \namenospace.
The test data sets are generated from the model
\begin{align*}
	Y ~=~ \alpha +\sum_{j=1}^p  f_j(X_j) + \sum_{k=1}^q \beta_k Z_k + \epsilon,\quad \epsilon\sim N(0,\sigma^2),
\end{align*}
where parameters are set to $\alpha=1$, $\beta_1=0.25$, $\beta_2=0.5$, $\beta_3=0.75$, $\beta_4=\dots=\beta_q=0$, and $\sigma^2=1$, and smooth functions are chosen as
\begin{align*}
	f_1(x)&=4x^2,\\
	f_2(x)&=\sin(2\pi x),\\
	f_3(x)&=3\exp\{-200(x-0.2)^2\}+0.5\exp\{-50(x-0.6)^2\},\\
	f_4(x)&=x,\\
	f_5(x)&=1.5x,\\
	f_6(x)&=2x,\\
	f_j(x)&=0, ~j=7,\dots,p,
\end{align*}
that is, three nonlinear functions and three linear functions are actually used for the smooth functions. (Note that in the data generating process, we need not consider the identifiability restriction $E[f_j(X_j)]=0$ since the functions are automatically adjusted by the intercept term.) Thus we have a $p$-dimensional vector $\bgam^{\rm x}=(2,2,2,1,1,1,0,\dots,0)^\top$ and a $q$-dimensional vector $\bgam^{\rm z}=(1,1,1,0,\dots,0)^\top$. 
The predictors $\{X_j,~j=1,\dots, p\}$ are generated by the Gaussian copula such that $X_j$ are marginally distributed uniformly on $(0,1)$ with pairwise correlation coefficient denoted by $\rho$.
The predictors $\{Z_k,~k=1,\dots,q\}$ are generated from a $q$-dimensional multivariate normal distribution with zero mean, identity covariance matrix, and pairwise correlation coefficient denoted by $\rho$.
We consider two different values for the pairwise correlation coeffiient ($\rho=0.5$ and $\rho=0.95$), and three different cases for the dimension of covariates ($p=q=10$, $p=q=25$, and $p=q=50$); note that a larger model with bigger $p$ and $q$ has more irrelevant variables because the number of nonzero coefficients are the same in all cases.
We also consider three different values for the sample size ($n=100$, $n=200$, and $n=500$), so that a total of 18 simulation settings are considered.
For each of the simulation settings, 100 independent data sets are generated to account for a sampling error.

\begin{table*}[t!]
	\caption{Simulation results with $p=q=10$. The performance measures are averaged over 100 replications.}
	\bigskip
	\centering
	\begin{tabular}{lllcccccccccc}
		\hline
		\noalign{\smallskip}
		&&& \multicolumn{4}{c}{Nonlinear components} & & \multicolumn{3}{c}{Linear components}\\
		\noalign{\smallskip}
		\cline{4-7} \cline{9-11}
		\noalign{\smallskip}
		$\rho$ & $n$ & Method & ${\rm MR_2^{\rm x}}$ & ${\rm MR_1^{\rm x}}$ & ${\rm MR_0^{\rm x}}$ & $\rm MR^{\rm x}$ && ${\rm MR_1^{\rm z}}$ & ${\rm MR_0^{\rm z}}$ & ${\rm MR^{\rm z}}$ && $\rm MR_T$\\
		\noalign{\smallskip}\hline\noalign{\smallskip}
		0.5 & 100 & \namenospace & 0.50 & 0.47 & 0.03 & 0.30 & & 0.27 & 0.05 & 0.12 & & 0.21\\
		& & Saban{\'e}s Bov{\'e} & 0.81 & 0.38 & 0.09 & 0.39 & & 0.32 & 0.07 & 0.14 & & 0.27\\
		& & Scheipl & 0.66 & 0.56 & 0.02 & 0.37 & & 0.45 & 0.00 & 0.14 & & 0.26\smallskip\\
		& 200 & \namenospace & 0.05 & 0.19 & 0.01 & 0.08 & & 0.11 & 0.05 & 0.07 & & 0.07\\
		& & Saban{\'e}s Bov{\'e} & 0.14 & 0.15 & 0.03 & 0.10 & & 0.14 & 0.04 & 0.07 & & 0.08\\
		& & Scheipl & 0.19 & 0.25 & 0.01 & 0.13 & & 0.28 & 0.00 & 0.08 & & 0.11\smallskip\\
		& 500 & \namenospace & 0.00 & 0.02 & 0.02 & 0.01 & & 0.01 & 0.02 & 0.02 & & 0.02\\
		& & Saban{\'e}s Bov{\'e} & 0.00 & 0.01 & 0.02 & 0.02 & & 0.02 & 0.02 & 0.02 & & 0.02\\
		& & Scheipl & 0.05 & 0.05 & 0.00 & 0.03 & & 0.15 & 0.00 & 0.04 & & 0.04\smallskip\\
		0.95 & 100 &  \namenospace & 0.78 & 0.98 & 0.01 & 0.53 & & 0.83 & 0.03 & 0.27 & & 0.40\\
		& & Saban{\'e}s Bov{\'e} & 0.79 & 0.97 & 0.01 & 0.53 & & 0.81 & 0.04 & 0.27 & & 0.40\\
		& & Scheipl & 0.97 & 1.00 & 0.00 & 0.59 & & 0.96 & 0.01 & 0.29 & & 0.44\smallskip\\
		& 200 & \namenospace & 0.47 & 0.85 & 0.02 & 0.40 & & 0.68 & 0.02 & 0.22 & & 0.31\\
		& & Saban{\'e}s Bov{\'e} & 0.52 & 0.87 & 0.01 & 0.42 & & 0.71 & 0.03 & 0.23 & & 0.33\\
		& & Scheipl & 0.65 & 0.96 & 0.01 & 0.48 & & 0.81 & 0.00 & 0.25 & & 0.37\smallskip\\
		& 500 & \namenospace & 0.22 & 0.56 & 0.01 & 0.24 & & 0.39 & 0.02 & 0.14 & & 0.18\\
		& & Saban{\'e}s Bov{\'e} & 0.24 & 0.59 & 0.00 & 0.25 & & 0.40 & 0.02 & 0.13 & & 0.19\\
		& & Scheipl & 0.39 & 0.74 & 0.00 & 0.34 & & 0.52 & 0.00 & 0.16 & & 0.25\\
		\noalign{\smallskip}\hline
	\end{tabular}
	\label{table:1}
\end{table*}

\begin{table*}[t!]
	\caption{Simulation results with $p=q=25$. The performance measures are averaged over 100 replications.}
	\bigskip
	\centering
	\begin{tabular}{lllcccccccccc}
		\hline
		\noalign{\smallskip}
		&&& \multicolumn{4}{c}{Nonlinear components} & & \multicolumn{3}{c}{Linear components}\\
		\noalign{\smallskip}
		\cline{4-7} \cline{9-11}
		\noalign{\smallskip}
		$\rho$ & $n$ & Method & ${\rm MR_2^{\rm x}}$ & ${\rm MR_1^{\rm x}}$ & ${\rm MR_0^{\rm x}}$ & $\rm MR^{\rm x}$ && ${\rm MR_1^{\rm z}}$ & ${\rm MR_0^{\rm z}}$ & ${\rm MR^{\rm z}}$ && $\rm MR_T$\\
		\noalign{\smallskip}\hline\noalign{\smallskip}
		0.5 & 100 &  \namenospace & 0.55 & 0.61 & 0.01 & 0.15 & & 0.43 & 0.01 & 0.06 & & 0.11\\
		& & Saban{\'e}s Bov{\'e} & 0.91 & 0.44 & 0.15 & 0.28 & & 0.41 & 0.12 & 0.16 & & 0.22\\
		& & Scheipl & 0.78 & 0.70 & 0.00 & 0.18 & & 0.61 & 0.00 & 0.07 & & 0.13\smallskip\\
		& 200 & \namenospace & 0.04 & 0.24 & 0.00 & 0.04 & & 0.21 & 0.01 & 0.04 & & 0.04\\
		& & Saban{\'e}s Bov{\'e} & 0.32 & 0.23 & 0.01 & 0.08 & & 0.25 & 0.01 & 0.04 & & 0.06\\
		& & Scheipl & 0.29 & 0.32 & 0.00 & 0.08 & & 0.35 & 0.00 & 0.04 & & 0.06\smallskip\\
		& 500 & \namenospace & 0.00 & 0.03 & 0.01 & 0.01 & & 0.06 & 0.00 & 0.01 & & 0.01\\
		& & Saban{\'e}s Bov{\'e} & 0.00 & 0.02 & 0.01 & 0.01 & & 0.06 & 0.00 & 0.01 & & 0.01\\
		& & Scheipl & 0.19 & 0.15 & 0.00 & 0.04 & & 0.20 & 0.00 & 0.02 & & 0.03\smallskip\\
		0.95 & 100 &  \namenospace & 0.85 & 0.98 & 0.01 & 0.22 & & 0.93 & 0.01 & 0.12 & & 0.17\\
		& & Saban{\'e}s Bov{\'e} & 0.88 & 0.98 & 0.01 & 0.23 & & 0.91 & 0.02 & 0.13 & & 0.18\\
		& & Scheipl & 0.99 & 1.00 & 0.00 & 0.24 & & 1.00 & 0.00 & 0.12 & & 0.18\smallskip\\
		& 200 & \namenospace & 0.55 & 0.95 & 0.00 & 0.18 & & 0.80 & 0.01 & 0.11 & & 0.14\\
		& & Saban{\'e}s Bov{\'e} & 0.62 & 0.92 & 0.00 & 0.19 & & 0.77 & 0.02 & 0.11 & & 0.15\\
		& & Scheipl & 0.82 & 1.00 & 0.00 & 0.22 & & 0.96 & 0.00 & 0.12 & & 0.17\smallskip\\
		& 500 & \namenospace & 0.22 & 0.65 & 0.01 & 0.11 & & 0.53 & 0.00 & 0.07 & & 0.09\\
		& & Saban{\'e}s Bov{\'e} & 0.23 & 0.65 & 0.00 & 0.11 & & 0.52 & 0.01 & 0.07 & & 0.09\\
		& & Scheipl & 0.46 & 0.88 & 0.00 & 0.16 & & 0.73 & 0.00 & 0.09 & & 0.12\\
		\noalign{\smallskip}\hline
	\end{tabular}
	\label{table:2}
\end{table*}

\begin{table*}[t!]
	\caption{Simulation results with $p=q=50$. The performance measures are averaged over 100 replications.}
	\bigskip
	\centering
	\begin{tabular}{lllcccccccccc}
		\hline
		\noalign{\smallskip}
		&&& \multicolumn{4}{c}{Nonlinear components} & & \multicolumn{3}{c}{Linear components}\\
		\noalign{\smallskip}
		\cline{4-7} \cline{9-11}
		\noalign{\smallskip}
		$\rho$ & $n$ & Method & ${\rm MR_2^{\rm x}}$ & ${\rm MR_1^{\rm x}}$ & ${\rm MR_0^{\rm x}}$ & $\rm MR^{\rm x}$ && ${\rm MR_1^{\rm z}}$ & ${\rm MR_0^{\rm z}}$ & ${\rm MR^{\rm z}}$ && $\rm MR_T$\\
		\noalign{\smallskip}\hline\noalign{\smallskip}
		0.5 & 100 &  \namenospace & 0.65 & 0.72 &  0.00 & 0.09 & & 0.52 & 0.01 & 0.04 & & 0.06\\
		& & Saban{\'e}s Bov{\'e} & - & - & - & - & & - & - & - & & - \\
		& & Scheipl & 0.84 & 0.80 & 0.00 & 0.10 & & 0.62 & 0.00 & 0.04 & & 0.07\smallskip\\
		& 200 & \namenospace & 0.06 & 0.28 &  0.00 & 0.02 & & 0.24 & 0.00 & 0.02 & & 0.02\\
		& & Saban{\'e}s Bov{\'e} & 0.90 & 0.21 & 0.24 & 0.28 & & 0.25 & 0.19 & 0.19 & & 0.24 \\
		& & Scheipl & 0.33 & 0.43 & 0.00 & 0.05 & & 0.37 & 0.00 & 0.02 & & 0.03\smallskip\\
		& 500 & \namenospace & 0.00 & 0.05 & 0.00 & 0.00 & & 0.07 & 0.00 & 0.01 & & 0.00\\
		& & Saban{\'e}s Bov{\'e} & 0.03 & 0.05 & 0.00 & 0.01 & & 0.08 & 0.00 & 0.01 & & 0.01 \\
		& & Scheipl & 0.23 & 0.20 & 0.00 & 0.03 & & 0.25 & 0.00 & 0.01 & & 0.02\smallskip\\
		0.95 & 100 &  \namenospace & 0.87 & 1.00 & 0.00 & 0.11 & & 0.92 & 0.01 & 0.06 & & 0.09 \\
		& & Saban{\'e}s Bov{\'e} & - & - & - & - & & - & - & - & & - \\
		& & Scheipl & - & - & - & - & & - & - & - & & -\smallskip\\
		& 200 & \namenospace & 0.60 & 0.95 & 0.00 & 0.10 & & 0.85 & 0.00 & 0.06 & & 0.08\\
		& & Saban{\'e}s Bov{\'e} & 0.82 & 0.95 & 0.01 & 0.11 & & 0.83 & 0.01 & 0.06 & & 0.09 \\
		& & Scheipl & 0.97 & 1.00 & 0.00 & 0.12 & & 0.99 & 0.00 & 0.06 & & 0.09\smallskip\\
		& 500 & \namenospace & 0.24 & 0.74 & 0.00 & 0.06 & & 0.59 & 0.00 & 0.04 & & 0.05\\
		& & Saban{\'e}s Bov{\'e} & 0.46 & 0.82 & 0.00 & 0.08 & & 0.56 & 0.01 & 0.04 & & 0.06 \\
		& & Scheipl & 0.54 & 0.96 & 0.00 & 0.09 & & 0.94 & 0.00 & 0.06 & & 0.07\\
		\noalign{\smallskip}\hline
	\end{tabular}
	\label{table:3}
\end{table*}

The effects of predictor variables are estimated using the method proposed in Section~\ref{sec:bvs}, and the misclassification rates for each possible value of the latent variables, $\bgam^{\rm x}$ and $\bgam^{\rm z}$, are computed as
\begin{align*}
	{\rm MR_\ell^{\rm x}}=\frac{\sum_{j=1}^p\mathbbm{1}(\hat\gamma_j^{\rm x}\ne\ell, \gamma_j^{\rm x}=\ell)}{\sum_{j=1}^p\mathbbm{1}(\gamma_j^{\rm x}=\ell)},\quad\ell=0,1,2,~~~~~{\rm MR_\ell^{\rm z}}=\frac{\sum_{k=1}^q\mathbbm{1}(\hat\gamma_k^{\rm z}\ne\ell, \gamma_k^{\rm z}=\ell)}{\sum_{k=1}^q\mathbbm{1}(\gamma_k^{\rm z}=\ell)},\quad\ell=0,1.
\end{align*}
The overall misclassification rates for linear and nonlinear components are defined as
\begin{align*}
	{\rm MR^{\rm x}}=\frac{\sum_{j=1}^p\mathbbm{1}(\hat\gamma_j^{\rm x}\ne\gamma_j^{\rm x})}{p},~~~~~{\rm MR^{\rm z}}=\frac{\sum_{k=1}^q\mathbbm{1}(\hat\gamma_k^{\rm z}\ne\gamma_k^{\rm z})}{q},
\end{align*}
and the total misclassification rate, $\rm MR_T$, is defined as a weighted average of $\rm MR^{\rm x}$ and $\rm MR^{\rm z}$.
We compare our \name with the Bayesian methods of \citet{bove:etal:15} and \citet{sche:etal:11}, which we refer to as ``Saban{\'e}s Bov{\'e}'' and ``Scheipl''. 
These methods are implemented by using the R packages, \texttt{hypergsplines} and \texttt{spikeSlabGAM}, respectively. (The \texttt{hypergsplines} package is not currently maintained in the repositories; we installed it in an old version of R using the source files.) For all methods, 20 knots (knot-candidates for MSBS) are used to approximate smooth functions. 
Since all methods suitably penalize smooth functions in the procedure of model selection, the precise number of knots (or knot candidates) is not very important and it suffices to use a value that is not too small.
Interestingly, we found that Scheipl is not invariant to the scale of a response variable, so observations are standardized before the analysis.

The results are presented in Tables~\ref{table:1}--\ref{table:3}, which show the performance measures averaged over the 100 replications. 
We observe that Saban{\'e}s Bov{\'e} and Scheipl do not work for some setups with $p=q=50$, and therefore those settings were excluded from Table~\ref{table:3}.
It is clear that \name always outperforms the other methods in terms of ${\rm MR_2^{\rm x}}$, the misclassification rate showing how well a method finds a true nonlinear component. This implies that \name can better identify a complex structure in a relationship between a predictor and a response. Tables~\ref{table:1} and~\ref{table:2} show that Saban{\'e}s Bov{\'e} and Scheipl tend to be good at identifying a zero or linear component, especially with small sample size. This means that these methods generally prefer a structure that is simpler than the true structure for the predictor-response relationship. When it comes to the overall performance measured by ${\rm MR^{\rm x}}$, ${\rm MR^{\rm z}}$, and ${\rm MR_T}$, \name clearly outperforms or is at least comparable to the other two methods. If our interest is primarily in understanding the true unknown complex structure for the effect of a predictor on a response, \name performs very well.

\begin{figure}[t!]
	\centering{\includegraphics[width=6in]{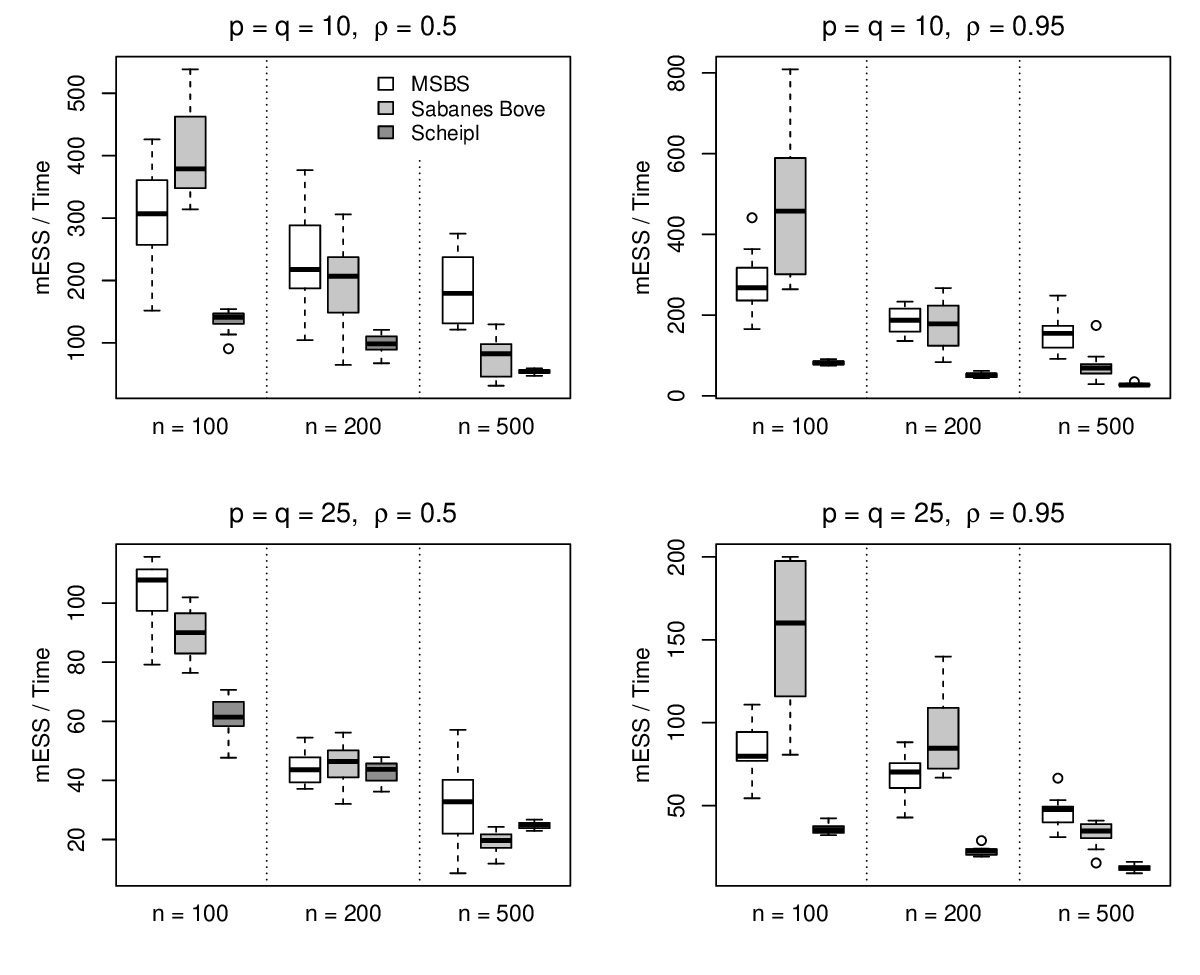}}
	\caption[]{Boxplots for multivariate effective sample size divided by runtime. The measures are averaged over 15 replications and show the computational advantage of \namenospace, especially for large sample sizes.}
	\label{fig:ess}
\end{figure}

Besides the misclassification rates, \name is compared with two other Bayesian methods in terms of computational efficiency, via the multivariate effective sample size (mESS) of the Markov chain \citep{vats:etal:15} divided by the runtime required to obtain 10,000 iterations after 1,000 burn-in iterations. This measure represents the number of independent draws of a Markov chain Monte Carlo algorithm per unit time.
The R package \texttt{mcmcse} gives an estimate of mESS.
To compute mESS, the posterior variance of each $\gamma_j^{\rm x}$ and of each $\gamma_k^{\rm z}$ needs to be sufficiently far from zero.
To achieve this, instead of using $\sigma^2=1$, here we use $\sigma^2=n/50$ and $n/100$ for the $\rho=0.5$ and $0.95$ cases, respectively. The case with $p=q=50$ is not considered for this simulation study, because too many MCMC iterations are required for the \texttt{mcmcse} package to work properly in this case. The performance measures are averaged over only 15 replications.
The comparison of computational efficiency is summarized in Figure~\ref{fig:ess}. In the small sample size case  ($n=100$), Saban{\'e}s Bov{\'e} tends to be more efficient than the other two methods, including \namenospace. As the sample size becomes larger ($n=200$ and $n=500$), however, the computational efficiency of \name significantly improves and  outperforms the other two methods.

The aim of the paper is to propose a new model selection strategy for additive partial linear models, Thus, we do not detail the performance of the function estimates proposed in Section~\ref{sec:est}. The advantages of the knot-selection technique in nonparametric regression, however, have been illustrated elsewhere \citep{smit:kohn:96,kohn:etal:01,jeon:park:16,jeon:etal:17,park:jeon:18}.

\section{Application to nutritional epidemiology study}
\label{sec:rd}

In this section, we consider the nutritional epidemiology study discussed by \citet{flet:fair:02} and \citet{liu:etal:11}. There is a close relationship between beta-carotene and certain types of cancer, such as lung, colon, breast, and prostate.
It is known that the antioxidant properties of beta-carotene help eliminate cancer-causing free radicals and thus can effectively reduce the risk of certain cancers. 
In addition, a sufficient beta-carotene supply can strengthen the body's autoimmune system to fight degenerative diseases such as cancer. Thus, it is of interest for clinicians and nutritionists to investigate the relationship between beta-carotene concentrations and other regulatory factors like age, gender, dietary intake, smoking status, and alcohol consumption. Some studies use simple linear models to explore the relationship, and the results are diverse and inconsistent \citep{nier:etal:89,faur:etal:06}. Therefore, there is a need for a more advanced statistical analysis to closely investigate the relationship between beta-carotene and cancer risk factors.

\name is thus applied to identify significant linear/nonlinear effects of risk factors on the logarithm of beta-carotene concentration.
The response variable is the plasma concentration of beta-carotene from 315
patients and the following risk factors are available: AGE (years), SEX (1=male,
2=female), SMOKSTAT (smoking status: 1=never, 2=former, 3=current Smoker), BMI (weight/height$^2$), VITUSE (vitamin use: 1=yes, fairly often, 2=yes, not often, 3=no), CALORIES (number
of calories consumed per day), FAT (grams of fat consumed per day), FIBER (grams of fiber consumed per day),
ALCOHOL (number of alcoholic drinks consumed per week), CHOL (mg of cholesterol consumed per day),
BETADIET (mcg of dietary beta-carotene consumed per day), and RETDIET (mcg of dietary retinol consumed per day).

\citet{liu:etal:11} used an additive partial linear model to examine the relationship between the logarithm of beta-carotene concentration and some personal characteristics by selecting significant linear predictors with their variable selection method after determining predictors having nonlinear effects via a preliminary analysis.  \citet{bane:ghos:14} used a similar model with more predictors and chose significant nonlinear predictors as well as linear predictors. The previous studies treated both AGE and CHOL as predictors with nonlinear effects. As can be seen in Figure~1 of \citet{liu:etal:11}, however, the significance of nonlinear effect of CHOL is uncertain. We thus apply \name to examine the true effects of the predictors on the logarithm of beta-carotene concentration.

\begin{table*}[t!]
	\caption{Results of model selection in the beta-carotene study. The columns show the posterior probabilities of the latent indicator variables for model selection and the estimates of the latent indicator variables under the best model.}
	\bigskip
	\centering
	\begin{tabular}{lrrrrrrrr}
		\hline\noalign{\smallskip}
		& \multicolumn{3}{c}{Posterior probability} \\
		\noalign{\smallskip}
		\cline{2-4}
		\noalign{\smallskip}
		Predictor & Zero & Linear & Nonlinear & & $\hat\bgam$\\
		\noalign{\smallskip}\hline\noalign{\smallskip}
		AGE & 0.034 & 0.017 & 0.949  & & 2\\
		SEX & 0.316 & 0.684 & - & & 1\\
		SMOKSTAT & 0.154 & 0.846 & - & & 1\\
		BMI & 0.000 & 0.899 & 0.101 & & 1\\
		VITUSE  & 0.053 & 0.947 & - & & 1\\
		CALORIES & 0.765 & 0.199 & 0.036  & & 0\\
		FAT   & 0.757 & 0.207 & 0.038  & & 0\\
		FIBER  & 0.104 & 0.732 & 0.164  & & 1\\
		ALCOHOL  & 0.836 & 0.119 & 0.045  & & 0  \\
		CHOL  & 0.622 & 0.252 & 0.126 & & 0   \\
		BETADIET  & 0.600 & 0.358 & 0.042 & & 0  \\
		RETDIET  & 0.844 & 0.120 & 0.036 & & 0  \\
		\noalign{\smallskip}\hline
	\end{tabular}
	\label{table:beta1}
\end{table*}

\begin{table*}[t!]
	\caption{Posterior summary statistics for predictors selected to have linear effects.}
	\bigskip
	\centering
	\begin{tabular}{lrr r}
		\hline\noalign{\smallskip}
		Parameter & Mean & Median & 95\% credible interval \\
		\noalign{\smallskip}\hline\noalign{\smallskip}
		Intercept & 5.462 & 5.462 & $(4.433,6.494)$ \\
		SEX & 0.290 & 0.289 & $(-0.054,0.635)$ \\
		SMOKSTAT & $-$0.141 & $-$0.141 & $(-0.303,0.019)$ \\
		BMI  & $-$0.033  & $-$0.033 & $(-0.052,-0.014)$ \\
		VITUSE  & $-$0.129 & $-$0.129 & $(-0.262, 0.003)$ \\
		FIBER & 0.024 & 0.024 & $(0.002,0.045)$  \\
		$\sigma^2$ & 0.411 & 0.410 & $(0.351, 0.482)$  \\
		\noalign{\smallskip}\hline
	\end{tabular}
	\label{table:beta2}
\end{table*}

\begin{figure}[t!]
	\centering{\includegraphics[width=3.5in]{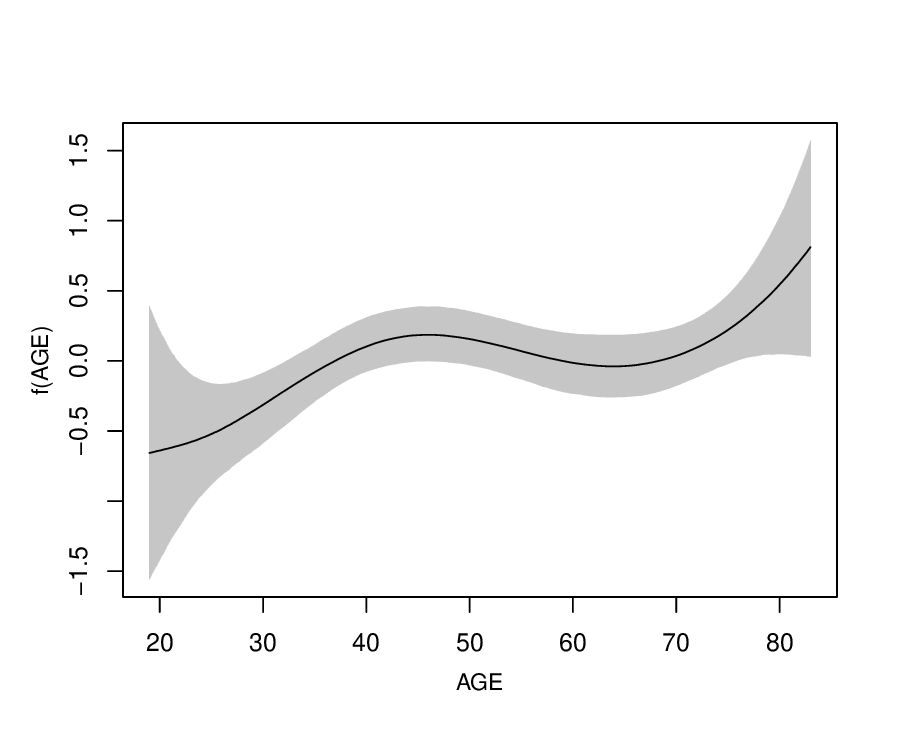}}
	\caption[]{Posterior estimates of a smooth function for AGE in the beta-carotene study. The solid line represents the point-wise posterior median of the function and the gray region represents the point-wise 95\% credible interval.}
	\label{fig:beta}
\end{figure}

We include the categorical variables SEX, SMOKSTAT, and VITUSE into the model as candidates of linear predictors, after centering in advance. The other covariates are continuous variables and are  candidates for nonlinear predictors. Table~\ref{table:beta1} shows the posterior probabilities of the latent indicator variables for model selection and the resulting estimate $\hat\bgam$. Our results are that \name selects SEX, SMOKSTAT, BMI, VITUSE, and FIBER as predictors with linear effects, and selects AGE as a predictor with a nonlinear effect. Interestingly, CHOL is chosen to have no significant effect on the logarithm of beta-carotene concentration in our analysis, which is the main difference between our analysis and former studies analyzing the same data. The posterior summary statistics for predictors selected to have linear effects are provided in Table~\ref{table:beta2}, and the estimated smooth function for AGE is given in Figure~\ref{fig:beta}.

\section{Discussion}
\label{sec:dc}
This paper proposes a new Bayesian model selection strategy and an efficient estimation method for additive partial linear models. Besides latent variables for model selection, we introduce additional latent variables that select basis functions to reduce a bias by data-adaptively controlling the smoothness of a nonlinear function. New prior distributions are carefully devised to overcome difficulties arising in the existing knot-selection methods. The resulting Bayesian model selection method, which we call \namenospace, outperforms the existing methods in terms of overall model selection accuracy and computational efficiency.

The main focus of this paper is on additive partial linear models with Gaussian errors, which can possibly provide a guide towards future research on the models with non-normal errors. Thus, \name developed in this paper can be generalized to analyze non-normal response data within a generalized linear model framework. Such a generalization can be based on recent studies about mixtures of $g$-priors for generalized linear models \citep{bove:held:11,li:clyd:15}.

It is also possible to naturally extend MSBS for a nonparametric additive model with multivariate additive components. The extension is straightforward if one uses a tensor product of splines \citep{debo:78} or radial basis functions in multiple dimensions \citep{kohn:etal:01}.
As another extension, one may consider an adaptive sampler for our model selection problem instead of the given Metropolis-Hastings algorithm. This may be achieved by using the currently available sampling schemes \citep{nott:kohn:05,ji:schm:13}.

\section*{Acknowledgement}
We thank the Associate Editor and the two anonymous referees for their constructive advice.
S. Jeong's research was supported by the Yonsei University Research Fund of 2021-22-0032.
T. Park's research was supported by Basic Science Research Program through the National Research Foundation of Korea(NRF) funded by the Ministry of Education (2017R1D1A1B03033536) and by the National Research Foundation of Korea(NRF) grant funded by the Korea government(MSIT) (2020R1A2C1A01005949).

\section*{Appendix}
\begin{appendices}
	
	\section{R package \texttt{MSBS}}
	\label{app:1}
	Here we show how to use the R package for \namenospace. Using the \texttt{devtools} package available at CRAN, our R package is installed and loaded by running
	\begin{verbatim}
		devtools::install_github("s-jeong/MSBS")
		library(MSBS)
	\end{verbatim}
	and then one can reproduce the results in Section~\ref{sec:rd} by running the examples in the help pages of the R functions \texttt{bmsaplm} and \texttt{estaplm}.

	\section{Proof of Proposition~\ref{thm:pr1}}
	\label{app:2}
	By writing $|\bgam^{\rm x}|=\sum_{j=1}^p\mathbbm{1}(\gamma_j^{\rm x}>0)$ and $|\bgam^{\rm z}|=\sum_{k=1}^q\gamma_k^{\rm z}$, we have
	\begin{align*}
		\pi(|\bgam|=k)&=\sum_{\ell=0\vee(k-q)}^{p\wedge k}\pi(|\bgam^{\rm x}|=\ell,|\bgam^{\rm z}|=k-\ell)=\sum_{\ell=0\vee(k-q)}^{p\wedge k}\sum_{\bgam:|\bgam^{\rm x}|=\ell,|\bgam^{\rm z}|=k-\ell}\pi(\bgam).
	\end{align*}
	Note that the cardinality of the set $\{\bgam:|\bgam^{\rm x}|=\ell,|\bgam^{\rm z}|=k-\ell\}$ for $\ell\in\{0\vee(k-q),\dots, p\wedge k\}$ is $\binom{p}{\ell}\binom{q}{k-\ell}2^\ell$ and $\pi(\bgam)$ is invariant on this set, where the factor $2^\ell$ appears because each $\gamma_j^{\rm x}$ has two nonzero values.
	Hence, the rightmost side of the last equation equals
	\begin{align*}
		\sum_{\ell=0\vee(k-q)}^{p\wedge k} \binom{p}{\ell}\binom{q}{k-\ell}B(k+1,p+q-k+1)=\binom{p+q}{k}B(k+1,p+q-k+1)=\frac{1}{p+q+1},
	\end{align*}
	which completes the proof.
	
	\section{Proof of Proposition~\ref{thm:pr2}}
	\label{app:3}
	Define $\bgam_{-j}^{\rm x}$ such that $\bgam^{\rm x}=\{\gamma_j^{\rm x},\bgam_{-j}^{\rm x}\}$ and let $|\bgam_{-j}^{\rm x}|=|\bgam^{\rm x}|-\mathbbm{1}(\gamma_j^{\rm x}>0)$. Using the same argument in the proof of Proposition~\ref{thm:pr1},
	\begin{align*}
		\pi(\gamma_j^{\rm x}=0)&=\sum_{k=0}^{p+q-1}\sum_{\ell=0\vee(k-q)}^{(p-1)\wedge k} \pi(\gamma_j^{\rm x}=0,|\bgam_{-j}^{\rm x}|=\ell,|\bgam^{\rm z}|=k-\ell)\\
		&=\sum_{k=0}^{p+q-1}\binom{p+q-1}{k}B(k+1,p+q-k+1)\\
		&=\sum_{k=0}^{p+q-1} \frac{p+q-k}{(p+q)(p+q+1)}=\frac{1}{2}.
	\end{align*}
	Because $\pi(\gamma_j^{\rm x}=1)$ and $\pi(\gamma_j^{\rm x}=2)$ are equal by the definition in~(\ref{eqn:bb}), it follows that $\pi(\gamma_j^{\rm x}=1)=\pi(\gamma_j^{\rm x}=2)=1/4$. Similarly, $\pi(\gamma_j^{\rm z}=0)=1/2$ can be verified.

\end{appendices}

\bibliographystyle{Chicago}

\bibliography{ref}

\begin{thebibliography}{}

\bibitem[\protect\citeauthoryear{Baladandayuthapani, Mallick, and
  Carroll}{Baladandayuthapani et~al.}{2005}]{bala:etal:2005}
Baladandayuthapani, V., B.~K. Mallick, and R.~J. Carroll (2005).
\newblock Spatially adaptive {B}ayesian penalized regression splines
  ({P}-splines).
\newblock {\em Journal of Computational and Graphical Statistics\/}~{\em
  14\/}(2), 378--394.

\bibitem[\protect\citeauthoryear{Banerjee and Ghosal}{Banerjee and
  Ghosal}{2014}]{bane:ghos:14}
Banerjee, S. and S.~Ghosal (2014).
\newblock Bayesian variable selection in generalized additive partial linear
  models.
\newblock {\em Stat\/}~{\em 3\/}(1), 363--378.

\bibitem[\protect\citeauthoryear{Barbieri and Berger}{Barbieri and
  Berger}{2004}]{barb:etal:04}
Barbieri, M.~M. and J.~O. Berger (2004).
\newblock Optimal predictive model selection.
\newblock {\em The Annals of Statistics\/}~{\em 32\/}(3), 870--897.

\bibitem[\protect\citeauthoryear{Bayarri, Berger, Forte, and
  Garc{\'\i}a-Donato}{Bayarri et~al.}{2012}]{baya:etal:12}
Bayarri, M.~J., J.~O. Berger, A.~Forte, and G.~Garc{\'\i}a-Donato (2012).
\newblock Criteria for {B}ayesian model choice with application to variable
  selection.
\newblock {\em The Annals of statistics\/}~{\em 40\/}(3), 1550--1577.

\bibitem[\protect\citeauthoryear{Belitser and Serra}{Belitser and
  Serra}{2014}]{beli:serr:14}
Belitser, E. and P.~Serra (2014).
\newblock Adaptive priors based on splines with random knots.
\newblock {\em Bayesian Analysis\/}~{\em 9\/}(4), 859--882.

\bibitem[\protect\citeauthoryear{Berger, Pericchi, and Varshavsky}{Berger
  et~al.}{1998}]{berg:etal:98}
Berger, J.~O., L.~R. Pericchi, and J.~A. Varshavsky (1998).
\newblock Bayes factors and marginal distributions in invariant situations.
\newblock {\em Sankhy{\=a}: The Indian Journal of Statistics, Series A\/}~{\em
  60\/}(3), 307--321.

\bibitem[\protect\citeauthoryear{Carlin and Chib}{Carlin and
  Chib}{1995}]{carl:chib:95}
Carlin, B.~P. and S.~Chib (1995).
\newblock Bayesian model choice via {M}arkov chain {M}onte {C}arlo methods.
\newblock {\em Journal of the Royal Statistical Society: Series B (Statistical
  Methodology)\/}~{\em 57\/}(3), 473--484.

\bibitem[\protect\citeauthoryear{Cripps, Carter, and Kohn}{Cripps
  et~al.}{2005}]{crip:etal:05}
Cripps, E., C.~Carter, and R.~Kohn (2005).
\newblock Variable selection and covariance selection in multivariate
  regression models.
\newblock {\em Handbook of Statistics\/}~{\em 25}, 519--552.

\bibitem[\protect\citeauthoryear{Curtis, Banerjee, and Ghosal}{Curtis
  et~al.}{2014}]{curt:etal:14}
Curtis, S.~M., S.~Banerjee, and S.~Ghosal (2014).
\newblock Fast {B}ayesian model assessment for nonparametric additive
  regression.
\newblock {\em Computational Statistics \& Data Analysis\/}~{\em 71}, 347--358.

\bibitem[\protect\citeauthoryear{de~Boor}{de~Boor}{1978}]{debo:78}
de~Boor, C. (1978).
\newblock {\em A Practical Guide to Splines}.
\newblock Springer.

\bibitem[\protect\citeauthoryear{Dellaportas, Forster, and
  Ntzoufras}{Dellaportas et~al.}{2002}]{dell:etal:02}
Dellaportas, P., J.~J. Forster, and I.~Ntzoufras (2002).
\newblock On {B}ayesian model and variable selection using {MCMC}.
\newblock {\em Statistics and Computing\/}~{\em 12\/}(1), 27--36.

\bibitem[\protect\citeauthoryear{Denison, Mallick, and Smith}{Denison
  et~al.}{1998}]{deni:etal:98}
Denison, D.~G., B.~K. Mallick, and A.~F. Smith (1998).
\newblock Bayesian {MARS}.
\newblock {\em Statistics and Computing\/}~{\em 8\/}(4), 337--346.

\bibitem[\protect\citeauthoryear{DiMatteo, Genovese, and Kass}{DiMatteo
  et~al.}{2001}]{dima:etal:01}
DiMatteo, I., C.~R. Genovese, and R.~E. Kass (2001).
\newblock Bayesian curve-fitting with free-knot splines.
\newblock {\em Biometrika\/}~{\em 88\/}(4), 1055--1071.

\bibitem[\protect\citeauthoryear{Faure, Preziosi, Roussel, Bertrais, Galan,
  Hercberg, and Favier}{Faure et~al.}{2006}]{faur:etal:06}
Faure, H., P.~Preziosi, A.~Roussel, S.~Bertrais, P.~Galan, S.~Hercberg, and
  A.~Favier (2006).
\newblock Factors influencing blood concentration of retinol,
  $\alpha$-tocopherol, vitamin c, and $\beta$-carotene in the french
  participants of the su. vi. max trial.
\newblock {\em European Journal of Clinical Nutrition\/}~{\em 60\/}(6),
  706--717.

\bibitem[\protect\citeauthoryear{Fernandez, Ley, and Steel}{Fernandez
  et~al.}{2001}]{fern:etal:01}
Fernandez, C., E.~Ley, and M.~F. Steel (2001).
\newblock Benchmark priors for {B}ayesian model averaging.
\newblock {\em Journal of Econometrics\/}~{\em 100\/}(2), 381--427.

\bibitem[\protect\citeauthoryear{Fletcher and Fairfield}{Fletcher and
  Fairfield}{2002}]{flet:fair:02}
Fletcher, R.~H. and K.~M. Fairfield (2002).
\newblock Vitamins for chronic disease prevention in adults: clinical
  applications.
\newblock {\em Journal of the American Medical Association\/}~{\em 287\/}(23),
  3127--3129.

\bibitem[\protect\citeauthoryear{George and Foster}{George and
  Foster}{2000}]{geor:fost:00}
George, E.~I. and D.~P. Foster (2000).
\newblock Calibration and empirical {B}ayes variable selection.
\newblock {\em Biometrika\/}~{\em 87\/}(4), 731--747.

\bibitem[\protect\citeauthoryear{Hobert and Casella}{Hobert and
  Casella}{1998}]{hobe:case:98}
Hobert, J.~P. and G.~Casella (1998).
\newblock Functional compatibility, {M}arkov chains, and {G}ibbs sampling with
  improper posteriors.
\newblock {\em Journal of Computational and Graphical Statistics\/}~{\em
  7\/}(1), 42--60.

\bibitem[\protect\citeauthoryear{Huang, Horowitz, and Wei}{Huang
  et~al.}{2010}]{huan:etal:10}
Huang, J., J.~L. Horowitz, and F.~Wei (2010).
\newblock Variable selection in nonparametric additive models.
\newblock {\em The Annals of Statistics\/}~{\em 38\/}(4), 2282--2313.

\bibitem[\protect\citeauthoryear{Jeong, Park, and Park}{Jeong
  et~al.}{2017}]{jeon:etal:17}
Jeong, S., M.~Park, and T.~Park (2017).
\newblock Analysis of binary longitudinal data with time-varying effects.
\newblock {\em Computational Statistics and Data Analysis\/}~{\em 112},
  145--153.

\bibitem[\protect\citeauthoryear{Jeong and Park}{Jeong and
  Park}{2016}]{jeon:park:16}
Jeong, S. and T.~Park (2016).
\newblock Bayesian semiparametric inference on functional relationships in
  linear mixed models.
\newblock {\em Bayesian Analysis\/}~{\em 11\/}(4), 1137--1163.

\bibitem[\protect\citeauthoryear{Ji and Schmidler}{Ji and
  Schmidler}{2013}]{ji:schm:13}
Ji, C. and S.~C. Schmidler (2013).
\newblock Adaptive {M}arkov chain {M}onte {C}arlo for {B}ayesian variable
  selection.
\newblock {\em Journal of Computational and Graphical Statistics\/}~{\em
  22\/}(3), 708--728.

\bibitem[\protect\citeauthoryear{Jullion and Lambert}{Jullion and
  Lambert}{2007}]{jull:lamb:07}
Jullion, A. and P.~Lambert (2007).
\newblock Robust specification of the roughness penalty prior distribution in
  spatially adaptive {B}ayesian p-splines models.
\newblock {\em Computational statistics \& data analysis\/}~{\em 51\/}(5),
  2542--2558.

\bibitem[\protect\citeauthoryear{Kohn, Smith, and Chan}{Kohn
  et~al.}{2001}]{kohn:etal:01}
Kohn, R., M.~Smith, and D.~Chan (2001).
\newblock Nonparametric regression using linear combinations of basis
  functions.
\newblock {\em Statistics and Computing\/}~{\em 11\/}(4), 313--322.

\bibitem[\protect\citeauthoryear{Lang and Brezger}{Lang and
  Brezger}{2004}]{lang:brez:04}
Lang, S. and A.~Brezger (2004).
\newblock Bayesian {P}-splines.
\newblock {\em Journal of computational and graphical statistics\/}~{\em
  13\/}(1), 183--212.

\bibitem[\protect\citeauthoryear{Li and Clyde}{Li and Clyde}{2018}]{li:clyd:15}
Li, Y. and M.~A. Clyde (2018).
\newblock Mixtures of $g$-priors in generalized linear models.
\newblock {\em Journal of the American Statistical Association\/}~{\em
  113\/}(524), 1828--1845.

\bibitem[\protect\citeauthoryear{Lian, Liang, and Ruppert}{Lian
  et~al.}{2015}]{lian:etal:15}
Lian, H., H.~Liang, and D.~Ruppert (2015).
\newblock Separation of covariates into nonparametric and parametric parts in
  high-dimensional partially linear additive models.
\newblock {\em Statistica Sinica\/}~{\em 25\/}(2), 591--607.

\bibitem[\protect\citeauthoryear{Liang, Paulo, Molina, Clyde, and Berger}{Liang
  et~al.}{2008}]{lian:etal:08}
Liang, F., R.~Paulo, G.~Molina, M.~A. Clyde, and J.~O. Berger (2008).
\newblock Mixtures of g priors for {B}ayesian variable selection.
\newblock {\em Journal of the American Statistical Association\/}~{\em
  103\/}(481), 410--423.

\bibitem[\protect\citeauthoryear{Lin and Zhang}{Lin and
  Zhang}{2006}]{lin:zhan:06}
Lin, Y. and H.~H. Zhang (2006).
\newblock Component selection and smoothing in multivariate nonparametric
  regression.
\newblock {\em The Annals of Statistics\/}~{\em 34\/}(5), 2272--2297.

\bibitem[\protect\citeauthoryear{Liu, Wang, and Liang}{Liu
  et~al.}{2011}]{liu:etal:11}
Liu, X., L.~Wang, and H.~Liang (2011).
\newblock Estimation and variable selection for semiparametric additive partial
  linear models.
\newblock {\em Statistica Sinica\/}~{\em 21\/}(3), 1225--1248.

\bibitem[\protect\citeauthoryear{Maruyama and George}{Maruyama and
  George}{2011}]{maru:geor:11}
Maruyama, Y. and E.~I. George (2011).
\newblock Fully {B}ayes factors with a generalized $g$-prior.
\newblock {\em The Annals of Statistics\/}~{\em 39\/}(5), 2740--2765.

\bibitem[\protect\citeauthoryear{Moreno, Gir{\'o}n, and Casella}{Moreno
  et~al.}{2015}]{more:etal:15}
Moreno, E., J.~Gir{\'o}n, and G.~Casella (2015).
\newblock Posterior model consistency in variable selection as the model
  dimension grows.
\newblock {\em Statistical Science\/}~{\em 30\/}(2), 228--241.

\bibitem[\protect\citeauthoryear{Nierenberg, Stukel, Baron, Dain, and
  Greenberg}{Nierenberg et~al.}{1989}]{nier:etal:89}
Nierenberg, D.~W., T.~A. Stukel, J.~A. Baron, B.~J. Dain, and E.~R. Greenberg
  (1989).
\newblock Determinants of plasma levels of beta-carotene and retinol.
\newblock {\em American Journal of Epidemiology\/}~{\em 130\/}(3), 511--521.

\bibitem[\protect\citeauthoryear{Nott and Kohn}{Nott and
  Kohn}{2005}]{nott:kohn:05}
Nott, D.~J. and R.~Kohn (2005).
\newblock Adaptive sampling for {B}ayesian variable selection.
\newblock {\em Biometrika\/}~{\em 92\/}(4), 747--763.

\bibitem[\protect\citeauthoryear{Park and Jeong}{Park and
  Jeong}{2018}]{park:jeon:18}
Park, T. and S.~Jeong (2018).
\newblock Analysis of {P}oisson varying-coefficient models with autoregression.
\newblock {\em Statistics\/}~{\em 52\/}(1), 34--49.

\bibitem[\protect\citeauthoryear{Park and Lee}{Park and
  Lee}{2021}]{park:lee:21}
Park, T. and S.~Lee (2021).
\newblock Improving the {G}ibbs sampler.
\newblock {\em Wiley Interdisciplinary Reviews: Computational Statistics\/},
  e1546.

\bibitem[\protect\citeauthoryear{Park and van Dyk}{Park and van
  Dyk}{2009}]{park:vand:09}
Park, T. and D.~A. van Dyk (2009).
\newblock Partially collapsed {G}ibbs samplers: Illustrations and applications.
\newblock {\em Journal of Computational and Graphical Statistics\/}~{\em
  18\/}(2), 283--305.

\bibitem[\protect\citeauthoryear{Raftery, Madigan, and Hoeting}{Raftery
  et~al.}{1997}]{raft:etal:97}
Raftery, A.~E., D.~Madigan, and J.~A. Hoeting (1997).
\newblock Bayesian model averaging for linear regression models.
\newblock {\em Journal of the American Statistical Association\/}~{\em
  92\/}(437), 179--191.

\bibitem[\protect\citeauthoryear{Raskutti, Wainwright, and Yu}{Raskutti
  et~al.}{2012}]{rask:etal:12}
Raskutti, G., M.~J. Wainwright, and B.~Yu (2012).
\newblock Minimax-optimal rates for sparse additive models over kernel classes
  via convex programming.
\newblock {\em Journal of Machine Learning Research\/}~{\em 13}, 389--427.

\bibitem[\protect\citeauthoryear{Ravikumar, Liu, Lafferty, and
  Wasserman}{Ravikumar et~al.}{2009}]{ravi:etal:09}
Ravikumar, P., H.~Liu, J.~Lafferty, and L.~Wasserman (2009).
\newblock Sparse additive models.
\newblock {\em Journal of the Royal Statistical Society. Series B (Statistical
  Methodology)\/}~{\em 71\/}(5), 1009--1030.

\bibitem[\protect\citeauthoryear{Rossell and Rubio}{Rossell and
  Rubio}{2019}]{ross:rubi:19}
Rossell, D. and F.~J. Rubio (2019).
\newblock Additive {B}ayesian variable selection under censoring and
  misspecification.
\newblock {\em arXiv preprint arXiv:1907.13563\/}.

\bibitem[\protect\citeauthoryear{Ruppert, Wand, and Carroll}{Ruppert
  et~al.}{2003}]{rupp:etal:03}
Ruppert, D., M.~P. Wand, and R.~J. Carroll (2003).
\newblock {\em Semiparametric Regression}.
\newblock Cambridge university press.

\bibitem[\protect\citeauthoryear{Saban{\'e}s~Bov{\'e} and
  Held}{Saban{\'e}s~Bov{\'e} and Held}{2011}]{bove:held:11}
Saban{\'e}s~Bov{\'e}, D. and L.~Held (2011).
\newblock Hyper-$ g $ priors for generalized linear models.
\newblock {\em Bayesian Analysis\/}~{\em 6\/}(3), 387--410.

\bibitem[\protect\citeauthoryear{Saban{\'e}s~Bov{\'e}, Held, and
  Kauermann}{Saban{\'e}s~Bov{\'e} et~al.}{2015}]{bove:etal:15}
Saban{\'e}s~Bov{\'e}, D., L.~Held, and G.~Kauermann (2015).
\newblock Objective {B}ayesian model selection in generalized additive models
  with penalized splines.
\newblock {\em Journal of Computational and Graphical Statistics\/}~{\em
  24\/}(2), 394--415.

\bibitem[\protect\citeauthoryear{Scheipl, Fahrmeir, and Kneib}{Scheipl
  et~al.}{2012}]{sche:etal:11}
Scheipl, F., L.~Fahrmeir, and T.~Kneib (2012).
\newblock Spike-and-slab priors for function selection in structured additive
  regression models.
\newblock {\em Journal of the American Statistical Association\/}~{\em
  107\/}(500), 1518--1532.

\bibitem[\protect\citeauthoryear{Scheipl and Kneib}{Scheipl and
  Kneib}{2009}]{sche:knei:09}
Scheipl, F. and T.~Kneib (2009).
\newblock Locally adaptive {B}ayesian p-splines with a normal-exponential-gamma
  prior.
\newblock {\em Computational Statistics \& Data Analysis\/}~{\em 53\/}(10),
  3533--3552.

\bibitem[\protect\citeauthoryear{Scott and Berger}{Scott and
  Berger}{2010}]{scot:berg:10}
Scott, J.~G. and J.~O. Berger (2010).
\newblock {B}ayes and empirical-{B}ayes multiplicity adjustment in the
  variable-selection problem.
\newblock {\em The Annals of Statistics\/}~{\em 38\/}(5), 2587--2619.

\bibitem[\protect\citeauthoryear{Shen and Ghosal}{Shen and
  Ghosal}{2015}]{shen:ghos:15}
Shen, W. and S.~Ghosal (2015).
\newblock Adaptive {B}ayesian procedures using random series priors.
\newblock {\em Scandinavian Journal of Statistics\/}~{\em 42\/}(4), 1194--1213.

\bibitem[\protect\citeauthoryear{Smith and Kohn}{Smith and
  Kohn}{1996}]{smit:kohn:96}
Smith, M. and R.~Kohn (1996).
\newblock Nonparametric regression using {B}ayesian variable selection.
\newblock {\em Journal of Econometrics\/}~{\em 75\/}(2), 317--343.

\bibitem[\protect\citeauthoryear{van Dyk, DeGennaro, Stein, Jefferys, and von
  Hippel}{van Dyk et~al.}{2009}]{vand:etal:09}
van Dyk, D.~A., S.~DeGennaro, N.~Stein, W.~H. Jefferys, and T.~von Hippel
  (2009).
\newblock Statistical analysis of stellar evolution.
\newblock {\em The Annals of Applied Statistics\/}~{\em 3\/}(1), 117--143.

\bibitem[\protect\citeauthoryear{van Dyk and Jiao}{van Dyk and
  Jiao}{2015}]{vand:jiao:15}
van Dyk, D.~A. and X.~Jiao (2015).
\newblock Metropolis-{H}astings within partially collapsed {G}ibbs samplers.
\newblock {\em Journal of Computational and Graphical Statistics\/}~{\em
  24\/}(2), 301--327.

\bibitem[\protect\citeauthoryear{van Dyk and Park}{van Dyk and
  Park}{2008}]{vand:park:08}
van Dyk, D.~A. and T.~Park (2008).
\newblock Partially collapsed {G}ibbs samplers: Theory and methods.
\newblock {\em Journal of the American Statistical Association\/}~{\em
  193\/}(482), 790--796.

\bibitem[\protect\citeauthoryear{Vats, Flegal, and Jones}{Vats
  et~al.}{2019}]{vats:etal:15}
Vats, D., J.~M. Flegal, and G.~L. Jones (2019).
\newblock Multivariate output analysis for markov chain monte carlo.
\newblock {\em Biometrika\/}~{\em 106\/}(2), 321--337.

\bibitem[\protect\citeauthoryear{Wang and Maruyama}{Wang and
  Maruyama}{2018}]{wang:maru:17}
Wang, M. and Y.~Maruyama (2018).
\newblock Posterior consistency of g-prior for variable selection with a
  growing number of parameters.
\newblock {\em Journal of Statistical Planning and Inference\/}~{\em 196},
  19--29.

\bibitem[\protect\citeauthoryear{Wu and Stefanski}{Wu and
  Stefanski}{2015}]{wu:stef:15}
Wu, Y. and L.~A. Stefanski (2015).
\newblock Automatic structure recovery for additive models.
\newblock {\em Biometrika\/}~{\em 102\/}(2), 381--395.

\bibitem[\protect\citeauthoryear{Xue}{Xue}{2009}]{xue:09}
Xue, L. (2009).
\newblock Consistent variable selection in additive models.
\newblock {\em Statistica Sinica\/}~{\em 19}, 1281--1296.

\bibitem[\protect\citeauthoryear{Zellner}{Zellner}{1986}]{zell:86}
Zellner, A. (1986).
\newblock On assessing prior distributions and {B}ayesian regression analysis
  with g-prior distributions.
\newblock {\em Bayesian inference and decision techniques: Essays in Honor of
  Bruno De Finetti\/}~{\em 6}, 233--243.

\end{thebibliography}
\end{document}